\documentclass[preprint,aps,nofootinbib,superscriptaddress,showpacs]{revtex4}
\usepackage{graphicx}
\usepackage{epsfig}
\usepackage{bm}
\usepackage{amssymb}

\newcommand{\be}{\begin{eqnarray}}
\newcommand{\ee}{\end{eqnarray}}

\begin{document}

\preprint{ADP-13-09/T829}

\title{Effects of the density-dependent weak form factors on the neutrino reaction via neutral current for the nucleon in nuclear matter and $^{12}$C}

\author{Myung-Ki Cheoun} \thanks{cheoun@ssu.ac.kr}
\author{Ki-Seok Choi} \thanks{kiseokchoi@gmail.com}
\affiliation{Department of Physics,
Soongsil University, Seoul 156-743, Korea}

\author{K. S. Kim}\thanks{kyungsik@kau.ac.kr}
\affiliation{School of Liberal Arts and Science, Korea Aerospace
University, Koyang 412-791, Korea}

\author{Koichi Saito} \thanks{koichi.saito@rs.tus.ac.jp}
\affiliation{  Department of Physics, Faculty of Science and Technology, Tokyo University of Science, Noda 278-8510, Japan}

\author{Toshitaka Kajino} \thanks{kajino@nao.ac.jp}
\affiliation{  National Astronomical Observatory of Japan, 2-21-1 Osawa,
Mitaka, Tokyo 181-8588, Japan \\
and Department of Astronomy, Graduate School of Science, University of
Tokyo, Hongo 7-3-1, Bunkyo-ku, Tokyo 113-0033, Japan }

\author{Kazuo Tsushima} \thanks{kazuo.tsushima@gmail.com}
\affiliation{CSSM, School of Chemistry and Physics, University of Adelaide, Adelaide SA 5005, Australia}
\affiliation{International Institute of Physics (IIP),
Federal University of Rio Grande do Norte (UFRN), Natal/RN 59078-400, Brazil}

\author{Tomoyuki~Maruyama} \thanks{maruyama.tomoyuki@nihon-u.ac.jp}
\affiliation{College of Bioresource Sciences, Nihon University,
Fujisawa 252-8510, Japan}

\date{\today}
\begin{abstract}
The nucleon form factors in free space are usually thought to be modified when a nucleon is bound
in a nucleus or immersed in a nuclear medium. We investigate effects of the density-dependent axial and weak-vector form factors
on the electro-neutrino ($\nu_e$) and anti-electro-neutrino $({\bar \nu_e})$ reactions with incident energy $E_{\nu} \le$  80 MeV via neutral current (NC) for a nucleon
in a nuclear medium or $^{12}$C. For the density-dependent form factors, we exploit the quark-meson-coupling (QMC) model,
and apply them to the $\nu_e$ and ${\bar \nu_e}$ induced reactions by NC.
About 12 \% decrease of the total cross section by the $\nu_e$ reaction on the nucleon is obtained at
normal density, $\rho = \rho_0 \sim 0.15 {fm}^{-3} $, as well as about 18 \% reduction of the total ${\nu}_e$
cross section on $^{12}$C, by the modification of the weak form factors of the bound nucleon.

However, similarly to the charged current reaction, effects of the nucleon property change in the ${\bar \nu}_e$ reaction reduce significantly the cross sections
about 30 \% for the nucleon in matter and $^{12}$C cases.
Such a large asymmetry in the ${\bar \nu}_e$ cross sections is addressed to originate
from the different helicities of ${\bar \nu}_e$ and ${\nu}_e$.
\end{abstract}

\maketitle

\section{Introduction}
In the core-collapse supernova (SN) explosion,
neutrino ($\nu$) heating is known as one of the main mechanisms for the explosion leading to the so-called
$\nu$-driven explosion \cite{Fisc11}.
Neutrinos emitted from the neutrino sphere propagate the proto-neutron star (PNS). The first propagating region is a core part of the PNS, whose density is believed to be about a few times of the normal nuclear density $\rho_0 \sim 0.15 {fm}^{-3} $. During the propagation, neutrinos may interact with nucleons in dense nuclear matter through two different modes; neutral current (NC) and charged current (CC). The former mediated by $Z^0$ boson corresponds to the neutrino scattering, while the latter through $W^{\pm}$ bosons is the neutrino absorption with the emission of corresponding leptons.

After passing by the uniform density region, $\rho_{uni} \sim 6.85 \times 10^{-2} f m^{-3} = 1.14 \times 10^{14} g / cm^{3} \sim 0.5 \rho_o$, which density dissolves a crust of the PNS into the core comprising a uniform plasma of nucleons and leptons, neutrinos enter into the crust region of the PNS. This crust region is usually treated as two different parts, inner and outer crusts divided by the neutron drip density, $\rho_{drip} \sim 2.70 \times 10^{-4} fm^{-3} = 4.48 \times 10^{11} g / cm^{3}$. Beyond this density,
neutrons drip out of finite nuclei presumed to be embedded as lattice structures in the outer crust of the neutron star.

Some of the interesting phenomena in the $\nu$ propagation inside the PNS by a unique property of the neutrino come from the asymmetry between the neutrino scattering and absorption due to strong magnetic fields
in the magnetar. For instance, the pulsar kick~\cite{Maru11,Maru12} and the spin deceleration~\cite{Maru13} of the strongly magnetized neutron stars were shown to be closely associated with the asymmetry, according to detailed studies of the neutrino transport in dense matter by a relativistic mean field theory (RMF).

Outside of the PNS, emitted neutrinos interact also with the nuclei already produced by the s-process in the progenitor
and/or the r-process in the explosion. Around the Si layer, the neutrinos may initiate the so called neutrino-proton ($\nu p$) process \cite{Wana11}. Namely, the anti-neutrino (${\bar \nu}$) absorption in proton-rich environment may produce neutrons immediately captured by neutron-deficient nuclei, which affects the proton process (p-process) by the $(n,p)$ or $(n, \gamma)$ reactions. In the O-Ne-Mg layer, whose density is assumed to be about $\rho \sim 10^{3} g /cm^{3}$, neutrino-induced reactions might play an important role of producing some p-nuclei, which are
odd-odd neutron-deficient nuclei. For example, the cosmological origins of $^{180}$Ta and $^{138}$La are believed
to originate from the $\nu$-process~\cite{Ch10-2,Ch12}. Other light nuclei abundances are also closely associated
with the neutrino interactions in He-C layer~\cite{Yosh08}.

Of course, the nuclear density outside the PNS is not so dense compared to that of inside the PNS.
But, since the nucleons interacting with the neutrinos are strongly bound, properties of such a bound nucleon are
expected to be modified from those in free space. Therefore, if such a drastic change of nuclear density happens
to the neutrino propagation, it would be of practical importance to investigate such medium effects or bound nucleon property changes on the neutrino propagation in both
inside and outside of the PNS.

Moreover, recently, strong evidences for the modification
of the nucleon properties in a nuclear medium have been
reported from the proton electromagnetic (EM) form factors
measured in polarized $({\vec e}, e^{'} {\vec p})$ scattering on $^{16}$O~\cite{Malo00} and
$^{4}$He~\cite{Diet01,Stra03,Paol10,Mala11,Brooks:2011sa} at MAMI and Jefferson Lab,
and also from the study of neutron properties in a nuclear medium through polarized
$({\vec e}, e^{'} {\vec n})$ scattering on $^{4}$He in Ref.~\cite{Cloe09}.
Since the weak vector currents and EM currents form iso-vector (vector) current,
one may expect naturally the modification of the weak vector form factors in a similar way to the EM form factors.
In addition, the fact that the bound neutron in a nucleus is nearly stable while a free neutron decays via the weak interaction with the life time of about 880 s,
implies that the dominant, axial vector form factor or axial coupling constant $g_A$,
in a nuclear medium is also to be modified and different from that in free space.

Thus, it is quite meaningful to investigate the change of the neutrino-induced reactions due to the modification
of nucleon properties in a nuclear medium, in order to pin down the ambiguities inherent in
the nucleon and/or nuclear structure on the interpretation of various neutrino reactions in the cosmos.
For the study of the nuclear weak structure, one needs more refined nuclear models,
because the nuclear reaction by the emitted neutrino energy from the PNS, whose energy range is less than 100 MeV, is sensitive on the collective motion of inside nucleons.

In our previous paper, we studied the medium effects on the neutrino reaction by charged current~\cite{DD1}.
A large asymmetry between the neutrino and anti-neutrino reactions in a dense nuclear medium is predicted.
In this study, we focus on the NC reaction with the density-dependent weak form factors estimated in
the quark-meson-coupling model (QMC)~\cite{QMCboundff,QMChe3ff,QMCmatterff,QMCmatterga}.
The model has been successfully applied to
study the properties of hadrons in nuclear matter,
finite nuclei and hypernuclei~\cite{QMCfinite,QMChyp,QMCreview}. For more through understanding of the medium effects or the effects by the change of nucleon properties in a nuclear medium,
the $\nu_e$ and ${\bar \nu}_e$ reactions on the nucleon in nuclear matter \cite{Kim04} as well as in $^{12}$C are examined in detail. Nuclear structure for $^{12}$C is treated
by Quasi-particle RPA (QRPA)~\cite{ch10,ch10-2}.

This paper is organized as follows.
Sec.~II is devoted to explain the form factors used in this study. Detailed discussions regarding the form factors in dense matter and their numerical results are addressed in Appendix A and B. Numerical results for the neutrino reaction via NC
on the nucleon in nuclear matter and $^{12}$C are presented in Sec.~III. Summary and conclusions are given in Sec.~IV.

\section{Weak form factors in nuclear weak current}

By the standard electro-weak theory, the weak current operator $W^{\mu}$ used for the $\nu$-induced reaction takes a
$V^{\mu} - A^{\mu}$ current form  which has isoscalar and isovector parts for NC interaction~\cite{ch10}:
\begin{eqnarray}\label{eq:w-current}
W^{\mu}  &=&  V_{3}^{\mu} - A_{3}^{\mu} - 2 {\sin}^2
{\theta}_{W} J_{em}^{\mu} - { 1 \over 2} ( V_s^{\mu} - A_s^{\mu}), \\ \nonumber
& = & ( 1 - 2 {\sin}^2 {\theta}_{W} ) V_{3}^{\mu} -
A_{3}^{\mu} - 2 {\sin}^2 {\theta}_{W} V_{0}^{\mu} - { 1 \over 2} (
V_s^{\mu} - A_s^{\mu}),
\end{eqnarray}
with the Weinberg angle $\theta_W$. Here $J_{em}^{\mu} = V_3^{\mu} +
V_0^{\mu}$, and $V_3^{\mu}$ and $A_3^{\mu}$ are plus components of the isovector
$V_{i}^{\mu}$ and $A_{i}^{\mu}$ by the isospin rotation, {\it i.e.} $V_3^{\mu} = V_{ 1 + 2 i}^{\mu}$ and $A_3^{\mu}
= A_{ 1 + 2 i }^{\mu}$.
Strangeness contributions, which are isoscalar parts, could be
considered at $ - { 1 \over 2} (V_{s}^{\mu} -  A_{s}^{\mu})$. For the charged current (CC) interaction, only $V_{3}^{\mu} - A_{3}^{\mu}$ term is
involved, so that the CC
reaction is nearly independent of the strangeness content in a nucleon.
For the elastic scattering of polarized electrons on the nucleon,
$J^{\mu} = - 2 {\sin}^2 {\theta}_{W} J_{em}^{\mu} - { 1 \over 2}
V_s^{\mu} $ is exploited, while only $J^{\mu}_{em} = V_{3}^{\mu} + V_{0}^{\mu}$ is usually taken for the meson electro-production.

For a free nucleon, the weak current operator comprises the vector, the
axial vector and the pseudo scalar form factors, $F_i^V (Q^2)$,
$F_A (Q^2)$ and $F_P (Q^2)$:
\begin{equation}\label{eq:w-current-FF}
{W}^{\mu}=F_{1}^V (Q^2){\gamma}^{\mu}+ F_{2}^V (Q^2){\frac {i}
{2M_N}}{\sigma}^{\mu\nu}q_{\nu}  + F_A(Q^2) \gamma^{\mu} \gamma^5 +
{ F_P(Q^2) \over {2M}} q^{\mu} \gamma^5~.
\end{equation}
Here we take the scalar form factor in the vectorial part and the axial tensor form factor in the axial part to be zero,
because of the conservation of the vector current (CVC) and no existence of the second class
current, respectively. By the CVC hypothesis with the inclusion of the isoscalar strange quark
contributions $F_i^s (Q^2)$, the vector form factors for protons
and neutrons $F_{i}^{V,~p(n)} (Q^2)$ are expressed as~\cite{giusti1}:
\begin{eqnarray}\label{eq:FF}
F_{i}^{V,p(n) (NC) } &=& ({\frac 1 2} - 2 \sin^2 \theta_W )
F_i^{p(n)} ( Q^2) - {\frac 1 2} F_i^{n(p)}( Q^2) -{\frac 1 2}
F_i^s ( Q^2) ~,\\ \nonumber F_{i}^{V,p(n) (CC) } &=& ( F_i^{p} ( Q^2) -
 F_i^{n}( Q^2))~.
\end{eqnarray}
The axial form factor is given by
\begin{equation}\label{eq:GA}
F^{NC}_A (Q^2) ={\frac 1 2} (\mp g_A + g_A^s)/(1+Q^2/M_A^2)^2~,~
  F_A^{CC} (Q^2)  = - g_A / {( 1 + Q^2
/ M_A^2)}^2~~~,
\end{equation}
where $g_A$ and $M_A$ are the axial coupling constant and the
axial cut off mass, respectively. The sign, $-(+)$ comes from the isospin
dependence of the target proton (neutron), respectively~\cite{musolf}.
The axial form factor in Eq.~(4) is just negative
to the form factor elsewhere, for example, in Ref.~\cite{giusti1},
because we take the $+$ sign for the $F_A (Q^2) $ in Eq.~(2). Although the ambiguity from the strangeness content
in a nucleon still persists~\cite{Ch08}, the contribution to total cross section is less than 10 \%
even in the quasi-elastic region~\cite{Ch-08}. Therefore, we do not take the strangeness contribution into account in this work.

Before applying to the neutrino reaction, we need to figure out the change of nucleon properties in a nuclear medium,
such as the effective nucleon mass, the axial coupling constant, the weak form factors of the nucleon.
Those properties are calculated in the
quark-meson coupling (QMC) model~\cite{QMCboundff,QMChe3ff,QMCmatterff,QMCmatterga}.
The constituent quark mass in a hadron is generated by the quark condensate, $\langle {\bar q} q \rangle$, in vacuum, but the mass (or $\langle {\bar q} q \rangle$) in nuclear matter may be reduced from the value in vacuum
because of the condensed scalar ($\sigma$) field depending on the nuclear density $\rho$.
The decrease of the quark mass then leads to the variation of baryon internal structures at the quark level.
Such effect are considered self-consistently in the QMC model.
Detailed features of the form factors and their modifications in nuclear matter used in this study are summarized in Appendix A and B.

%
%
\begin{figure}
\centering
\includegraphics[width=8.0cm]{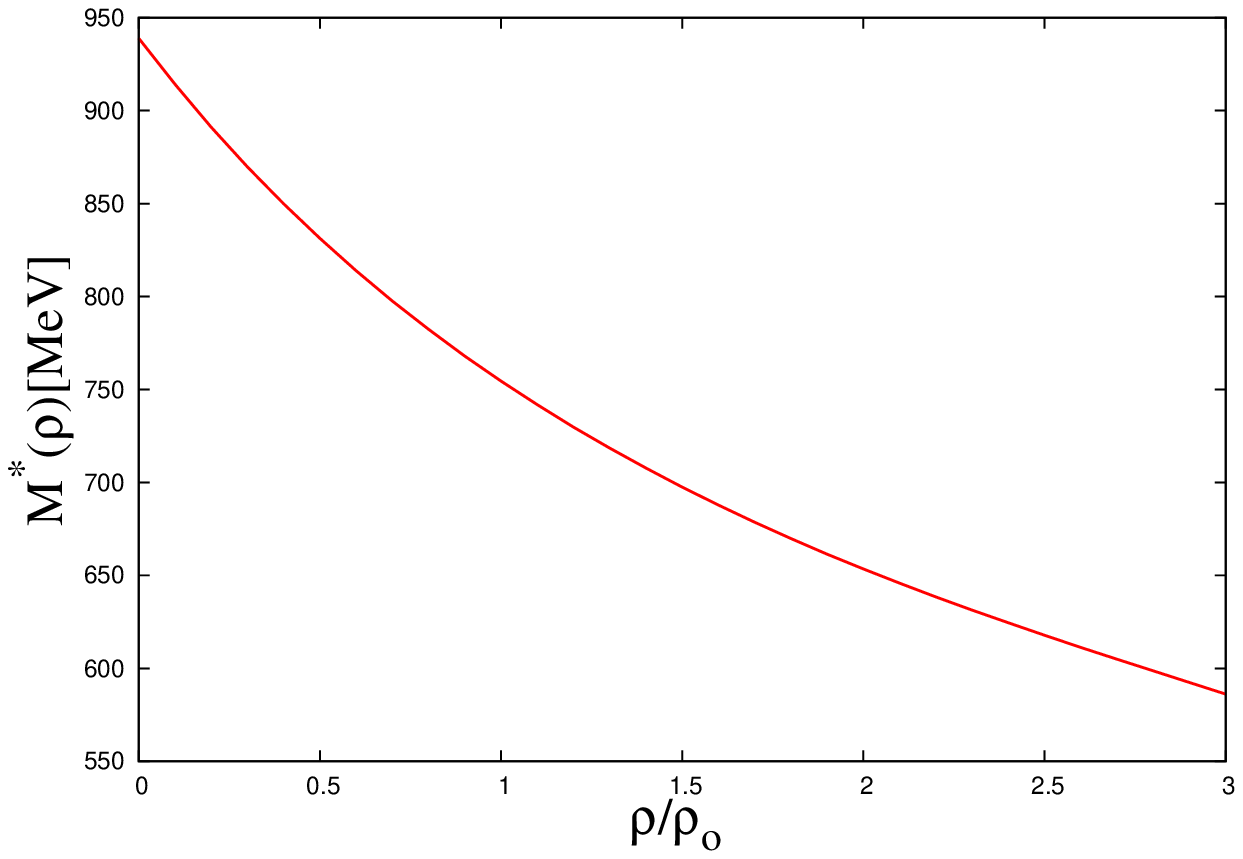}
\includegraphics[width=8.0cm]{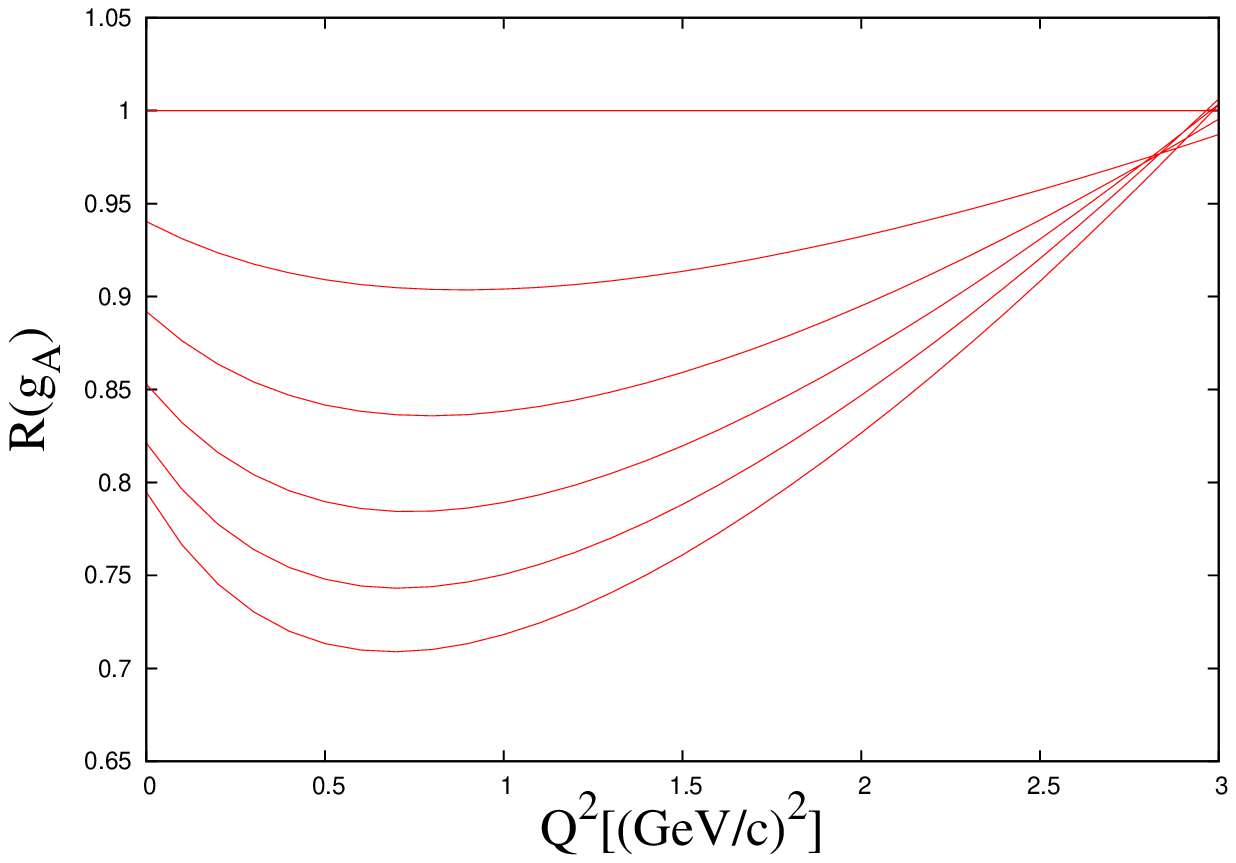}
\caption{(Color online) Effective nucleon mass $M^*(\rho)$ presented in terms of the nuclear density ratio
$\rho / \rho_o$ (left), and the axial vector form factor normalized to that in free space (right),
$R(g_A) = g_A(\rho, Q^2)/g_A(\rho = 0, Q^2)$, with finite momentum transfer in nuclear matter.
From the uppermost (vacuum) in the right panel, the density ratio is increased by 0.5 $\rho_o$.
The lowermost curve is for $\rho = 2.5 \rho_o$.} \label{fig1:massga}
\end{figure}

In Fig.~\ref{fig1:massga}, effective nucleon mass in nuclear matter, $M^* (\rho)$, is illustrated,
which shows a monotonic decrease with the increase of nuclear density.
The modification of the axial vector form factor in nuclear matter is also shown in the right panel in Fig.~1
as a function of four momentum transfer $Q^2 {[GeV/c]}^2$, which is normalized to that in free space,
$R(g_A) = g_A(\rho, Q^2)/g_A(\rho = 0, Q^2)$.
Even in the small momentum transfer region, where most of the neutrino reaction in the cosmos expected to occur,
the reduction of the axial form factor $g_A (\rho, Q^2)$ amounts to 11\% around $\rho = \rho_o$.
In the quasi-elastic region, the $Q^2$ dependence of the form factors on each density becomes more significant.
Detailed discussions on the change of vector form factors in a nuclear medium adopted in this paper
are summarized as figures in Appendix B.
\section{ Effects of density dependent weak form factors on the neutrino reaction via NC on the nucleon in nuclear matter and $^{12}$C}
\subsection{Results on the Nucleon in Dense Matter}
By using the following Sachs form factors and $G_A^{NC} = F_A^{NC}$
\begin{equation}\label{eq:Sachff}
G_E^V (Q^2) = F_1^V (Q^2) - { {Q^2} \over { 4 M^2}} F_2^V
(Q^2)~,G_M^V (Q^2) = F_1^V (Q^2) + F_2^V (Q^2)~,
\end{equation}
we calculated differential cross sections of the neutrino (antineutrino) reactions on the nucleon via NC as follows ~\cite{Ch08,Albe02}
\begin{eqnarray}\label{eq:Diffcc}
{( {  {d \sigma  } \over {d Q^2  }} )}_{\nu ({\bar \nu}   )
}^{NC}&=& { {G_F^2 } \over {2 \pi }} [ { 1 \over 2} y^2 { (G_M
)}^2 + ( 1 - y - { { M }\over {2 E_{\nu} }} y ) { {{ (G_E
)}^2 + { E_{\nu} \over 2 M} y { (G_M )}^2 } \over {1 + {  E_{\nu} \over 2 M} y }} \\
\nonumber & & + ( { 1 \over 2} y^2 + 1 - y + { { M } \over {2
E_{\nu} }} y ){ (G_A )}^2 \mp 2 y ( 1 - { 1 \over 2 } y ) G_M G_A
]~,\\ \nonumber {( {  {d \sigma  } \over {d Q^2  }} )}_{\nu ({\bar
\nu}   ) }^{CC}&=& {( {  {d \sigma  } \over {d Q^2  }} )}_{\nu
({\bar \nu}   ) }^{NC} ( G_E \rightarrow G_E^{CC}, G_M \rightarrow
G_M^{CC}, G_A \rightarrow G_A^{CC})~,
\end{eqnarray}
with
\begin{equation}\label{eq:Sachff2}
G_{E}^{CC} = G_E^p (Q^2) - G_E^n (Q^2)~,~G_{M}^{CC} = G_M^p (Q^2)
- G_M^n (Q^2)~.
\end{equation}
Here we omit superscript 'V' for the form factors. $E_{\nu}$ is the energy of the incident $\nu ({\bar \nu})$ in the
laboratory frame, and $y = { { p \cdot q } / { p \cdot k }} = { {
Q^2 } / {2 p \cdot k }}$ with $k,p$ and $q$ being respectively initial 4-momenta of
$\nu ({\bar \nu})$ and target nucleon, and 4-momentum transfer to
the nucleon. The sign, $-$ ($+$), corresponds to the
$\nu$ (${\bar \nu}$). Therefore, the difference and the sum of
the cross sections are simply summarized as
\begin{equation}\label{eq:Difmin}
{( {  {d \sigma  } \over {d Q^2  }} )}_{\nu     }^{NC} - {( { {d
\sigma  } \over {d Q^2  }} )}_{ {\bar \nu}    }^{NC} = - { {G_F^2
} \over { 2 \pi }}~ 4 y ( 1 - { 1 \over 2} y ) G_M G_A~,
\end{equation}
\begin{eqnarray}\label{eq:Difplu}
{( {  {d \sigma  } \over {d Q^2  }} )}_{\nu     }^{NC} + {( { {d
\sigma  } \over {d Q^2  }} )}_{ {\bar \nu}    }^{NC} &=& { {G_F^2
} \over { \pi }} [ { 1 \over 2} y^2 { (G_M )}^2 + ( 1 - y - { { M
}\over {2 E_{\nu} }} y ) { {{ (G_E
)}^2 + { E_{\nu} \over 2 M} y { (G_M )}^2 } \over {1 + {  E_{\nu} \over 2 M} y }} \\
\nonumber & & + ( { 1 \over 2} y^2 + 1 - y + { { M } \over {2
E_{\nu} }} y ){ (G_A )}^2 ]~,
\end{eqnarray}
and those via CC case are given by the replacements, $G_E \rightarrow
G_E^{CC}, G_M \rightarrow G_M^{CC}, G_A \rightarrow G_A^{CC} (= F_A^{CC})$ in Eqs.~(\ref{eq:Difmin})
and~(\ref{eq:Difplu}).

\begin{figure}
\centering
\includegraphics[width=8.0cm]{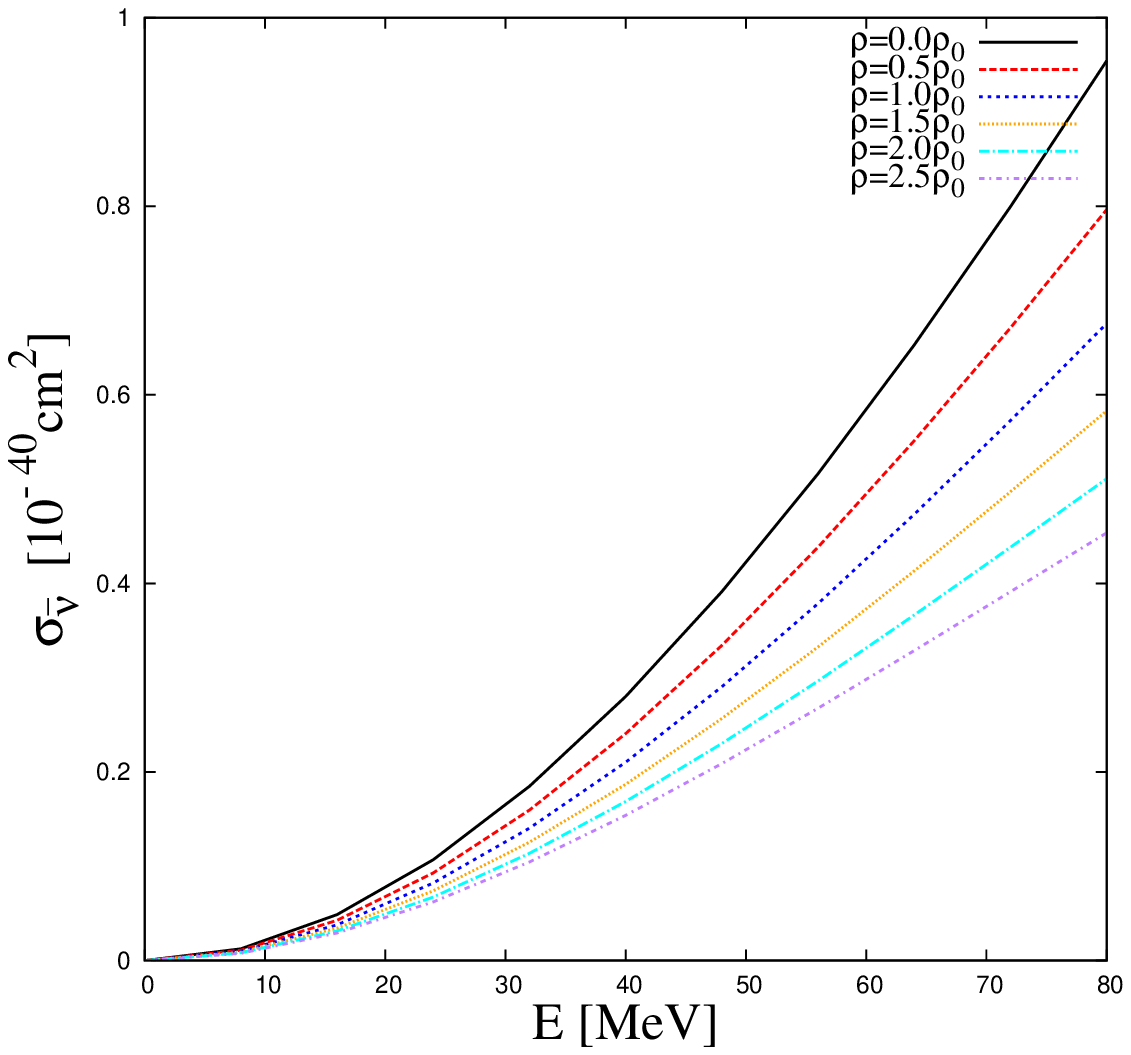}
\includegraphics[width=8.0cm]{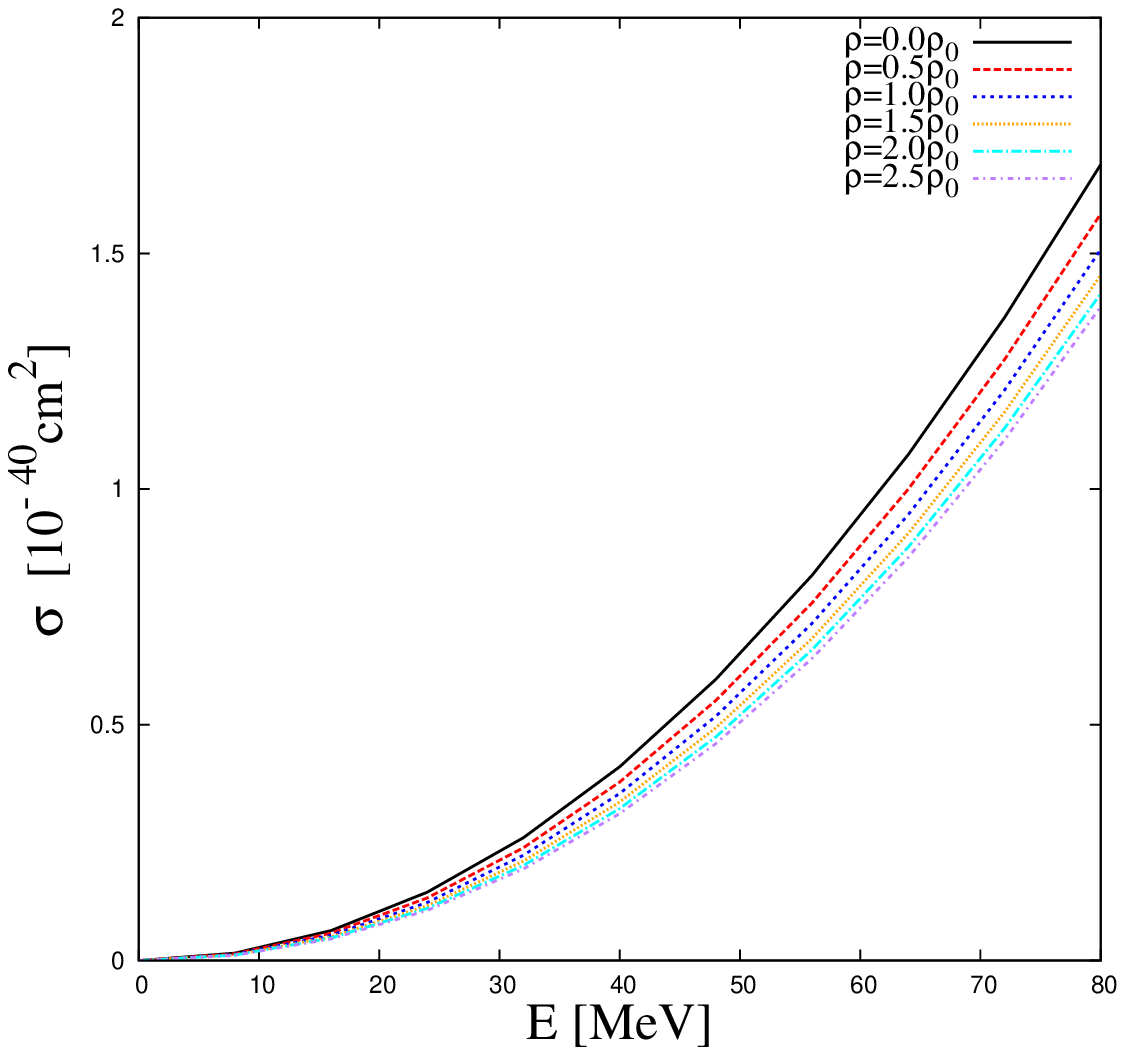}
\caption{(Color Online)  Density dependence of the total cross sections for ${\bar \nu_e} + p \rightarrow {\bar \nu_e}^{'} + p $ (left) and  ${ \nu_e} + n \rightarrow { \nu_e}^{'} + n  $ (right) in nuclear matter. The y axis is $10^{-40} cm^2$,
while the x-axis is the incident neutrino energy in MeV. Black (solid) curves are the results in free space.
Cross sections decrease with increasing the density, by $\rho / \rho_o$ = 0.5 (red (dashed)), 1.0 (blue (dotted)), 1.5 (yellow (short dotted)), 2.0 (sky-blue (dot long dashed)) and 2.5(cyan (dot short dashed)) in both reactions.} \label{fig2:nucleonNC}
\end{figure}
%

Numerical results obtained using the in-medium modified weak form factors are summarized in Fig.~\ref{fig2:nucleonNC},
where total cross sections for ${\bar \nu_e} + p \rightarrow {\bar \nu_e}^{'} + p $ (left) and ${ \nu_e} + n \rightarrow { \nu_e}^{'} + n  $ (right) via NC in nuclear matter are presented. Total cross sections decrease about 15\% per each density increase step until the
$\rho \sim 2.5 \rho_0$ (sky-blue (dot dashed) curve). However, for the ${ \nu_e} + n \rightarrow { \nu_e}^{'} + n $ reaction,
the variation in nuclear matter below $E_{\nu} \sim$ 30 MeV
is less than 3 \%. Even the cross sections around $E_{\nu} \sim$ 80 MeV decrease about 12 \% maximally at $\rho = \rho_0$
(blue (dotted) curve) compared to that in free space (black (solid) curve).

\begin{figure}
\centering
\includegraphics[width=8.0cm]{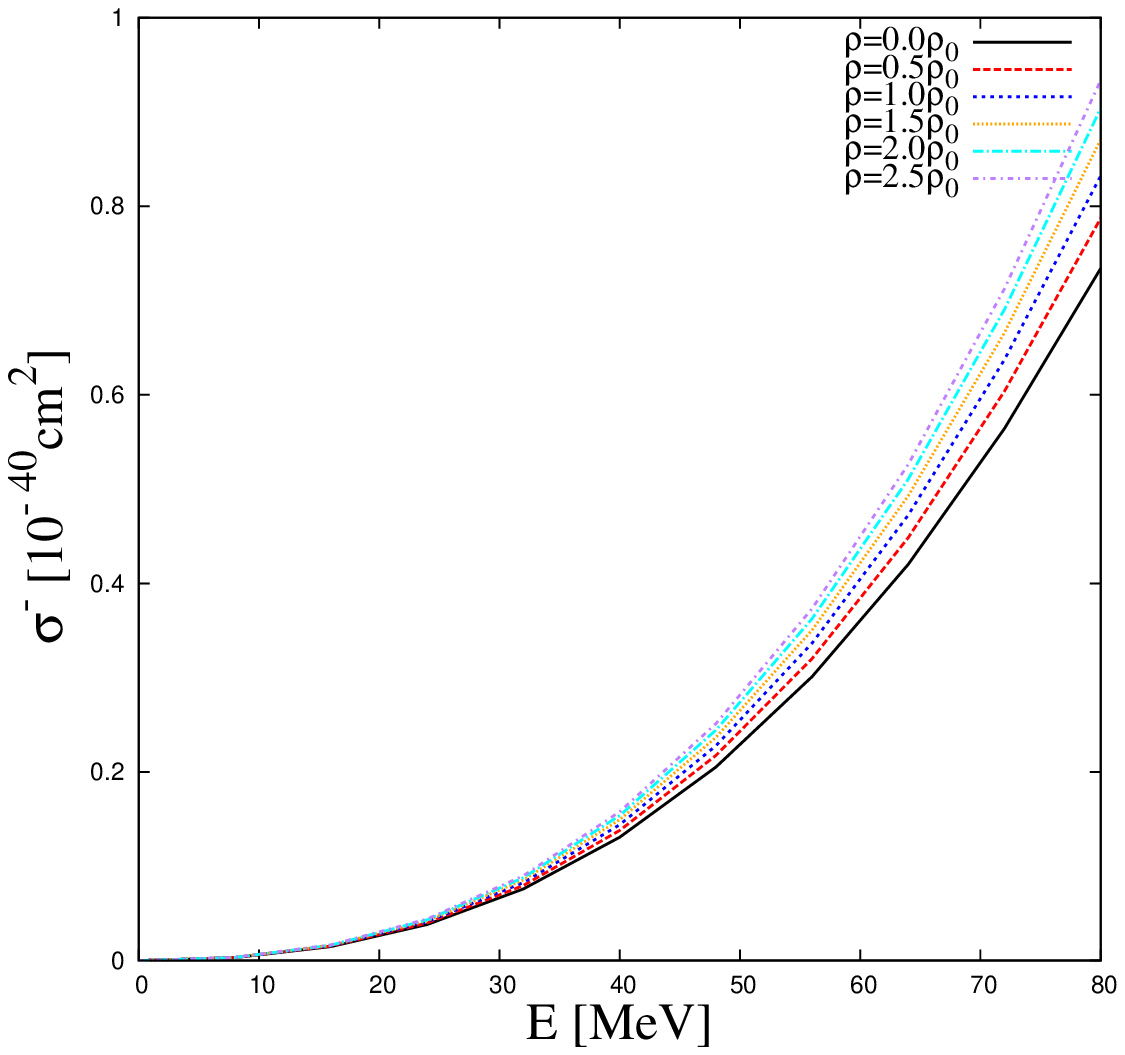}
\includegraphics[width=8.0cm]{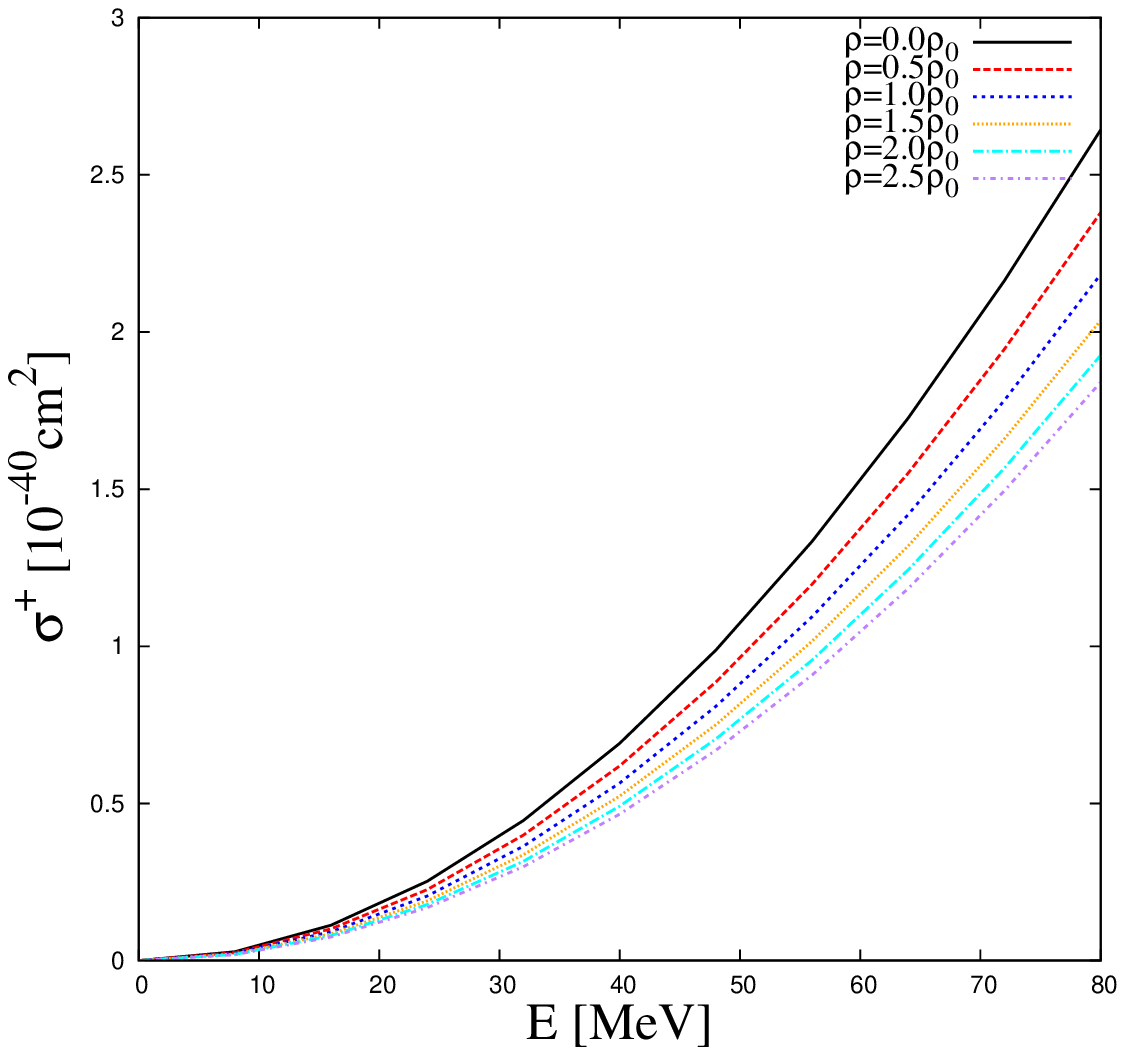}
\caption{(Color Online) Comparison of the total cross sections via the NC reaction for
$\sigma^- =\sigma (\nu_e) - \sigma ( {\bar \nu}_e)$ (left) and $\sigma^+ =\sigma (\nu_e) + \sigma ( {\bar \nu}_e)$ (right),
which respectively correspond to the HD and HID terms in Eq.~(\ref{eq:Diffcc}).
Legends for the curves are the same as those for Fig.~\ref{fig2:nucleonNC}.} \label{fig3:asymnucleonNC}
\end{figure}

This large asymmetry for the ${\bar \nu}_e$ and ${\nu}_e$ reactions due to the change in the nucleon properties in
a nuclear medium, which was also found in the CC reaction~\cite{DD1},
can be understood by the last, helicity-dependent (HD) term in
Eq.~(\ref{eq:Diffcc}) in an analogous way to the CC case \cite{DD1}.
The HD term contribution can be estimated from the asymmetry in the neutrino reaction,
$ \sigma^{-} = \sigma (\nu_e) - \sigma ( {\bar \nu}_e)$. For example, in Fig.~\ref{fig3:asymnucleonNC},
we plot the related cross sections, $\sigma^-$
and $\sigma^+ = \sigma (\nu_e) + \sigma ( {\bar \nu}_e)$, which respectively correspond to the HD and
helicity independent (HID) term in Eq.~(\ref{eq:Diffcc}). The HD term in the left panel shows the increase of
the cross sections with increasing the nuclear density, while the HID term in the right panel shows
the decrease of the cross sections with increasing the density.

For the ${\nu}_e$ reaction, which is a half of the sum of $\sigma^-$ and $\sigma^+$ shown in the both panels,
the HD term plays a countervailing role of the medium effects leading to the smaller effects,
while the HD term enhances the medium effects for the ${\bar \nu}_e$ reaction.
Therefore, the large asymmetry between the $\nu_e$ and ${\bar \nu}_e$ reaction cross section comes from the different
helicities of $\nu$ and ${\bar \nu}$. Radiative corrections in the ${\nu}_e$ reaction are not taken into account
in this work, because the effects are known to be less than 2 \%~\cite{Stru03}.
If we compare present NC results with those of the CC reaction in our preceding paper \cite{DD1}, this asymmetry can be arisen in the $\nu$ reaction, irrespective of the current types.
%

%
%
%
\subsection{Results on the Bound Nucleon in $^{12}$C}
To calculate the neutrino reaction on $^{12}$C, we use the following differential cross section formula,
whose detailed explanations can be found in Ref.~\cite{ch10},
\begin{eqnarray}\label{eq:DiffccC} & & ({{d \sigma_{\nu}} \over {d \Omega }  })_{(\nu / {\bar
\nu})} = { { G_F^2 \epsilon k ~  } \over {\pi ~ (2 J_i + 1 ) }}~
\bigl[ ~ {\mathop\Sigma_{J = 0}} (
 1+ {\vec \nu} \cdot {\vec \beta }){| <  J_f || {\hat {\cal M}}_J || J_i > | }^2
 \\ \nonumber & & + (
 1 - {\vec \nu} \cdot {\vec \beta } + 2({\hat \nu} \cdot {\hat q} )
 ({\hat q} \cdot {\vec \beta}  ))
  {| <  J_f || {\hat {\cal L}}_J ||
J_i > | }^2  -
 {\hat q} \cdot ({\hat \nu}+ {\vec \beta} )  { 2 Re < J_f || {\hat {\cal L}}_J  || J_i>
{< J_f|| {\hat {\cal M}}_J || J_i >}^*  } \\
\nonumber & &  + {\mathop\Sigma_{J = 1}} ( 1 - ({\hat \nu} \cdot
{\hat q} )({\hat q} \cdot {\vec \beta}  ) ) ( {| <  J_f || {\hat
{\cal T}}_J^{el}  || J_i > | }^2 + {| <  J_f || { {\hat
{\cal T}}}_J^{mag} || J_i > | }^2
) \\
\nonumber & &  \pm {\mathop\Sigma_{J = 1}} {\hat q} \cdot ({\hat
\nu} - {\vec \beta} )  2 Re [ <  J_f || { {\hat {\cal T}}}_J^{mag} ||
J_i > {<  J_f || {{\hat {\cal T}}}_J^{el} || J_i > }^* ]\bigr]~,
\end{eqnarray}
where ${\vec \nu}$ and $ {\vec k}$ are the 3-momenta of the incident and final
neutrinos, and ${\vec q} = {\vec k} - {\vec \nu}$, ${\vec \beta} =
{\vec k} / \epsilon $ with the final neutrino's energy $\epsilon$.
Of course, the extremely relativistic limit (ERL) may yield more
simple formula, but we use the general expression to get accurate results.
\begin{figure}
\centering
\includegraphics[width=8.0cm]{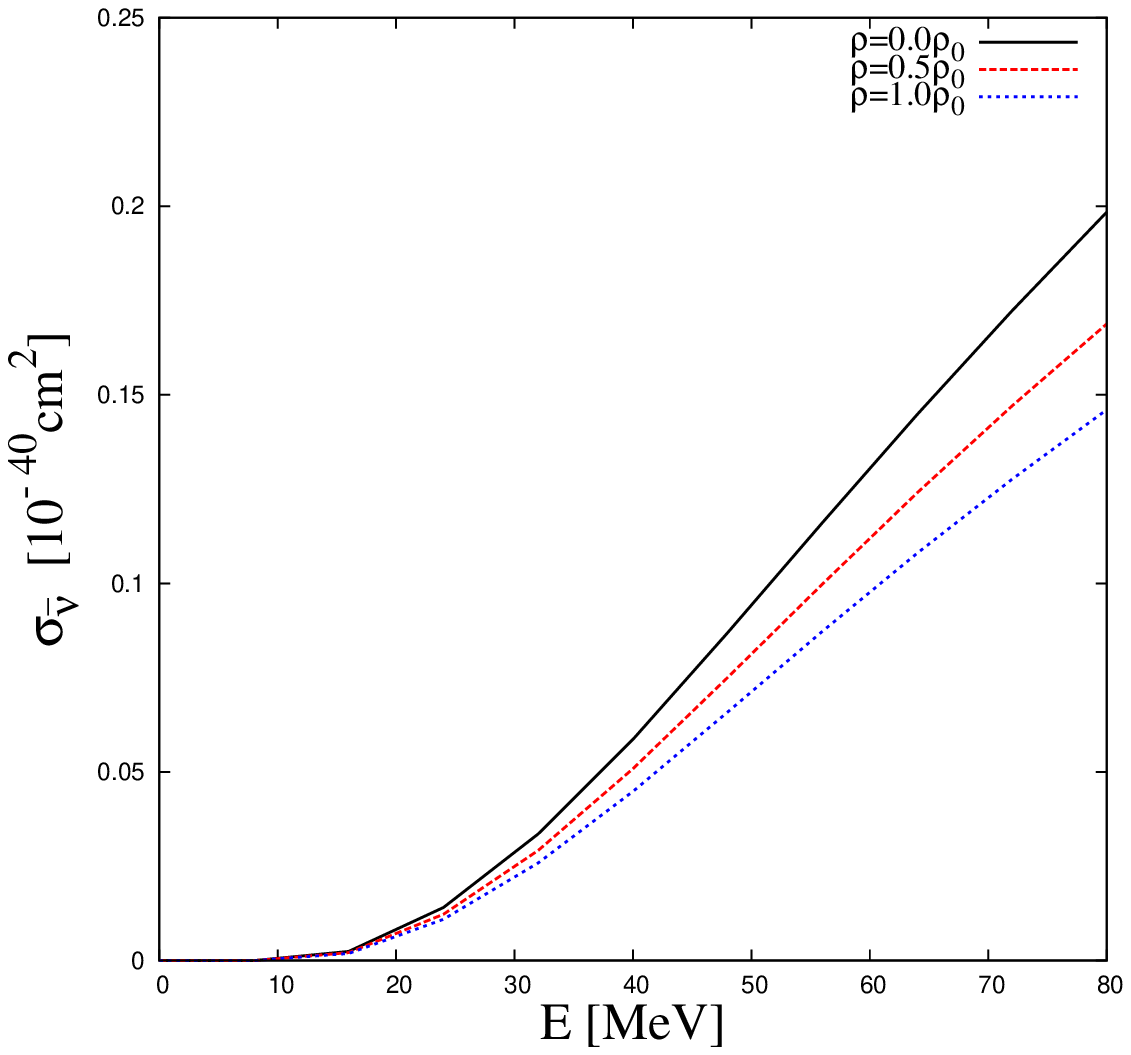}
\includegraphics[width=8.0cm]{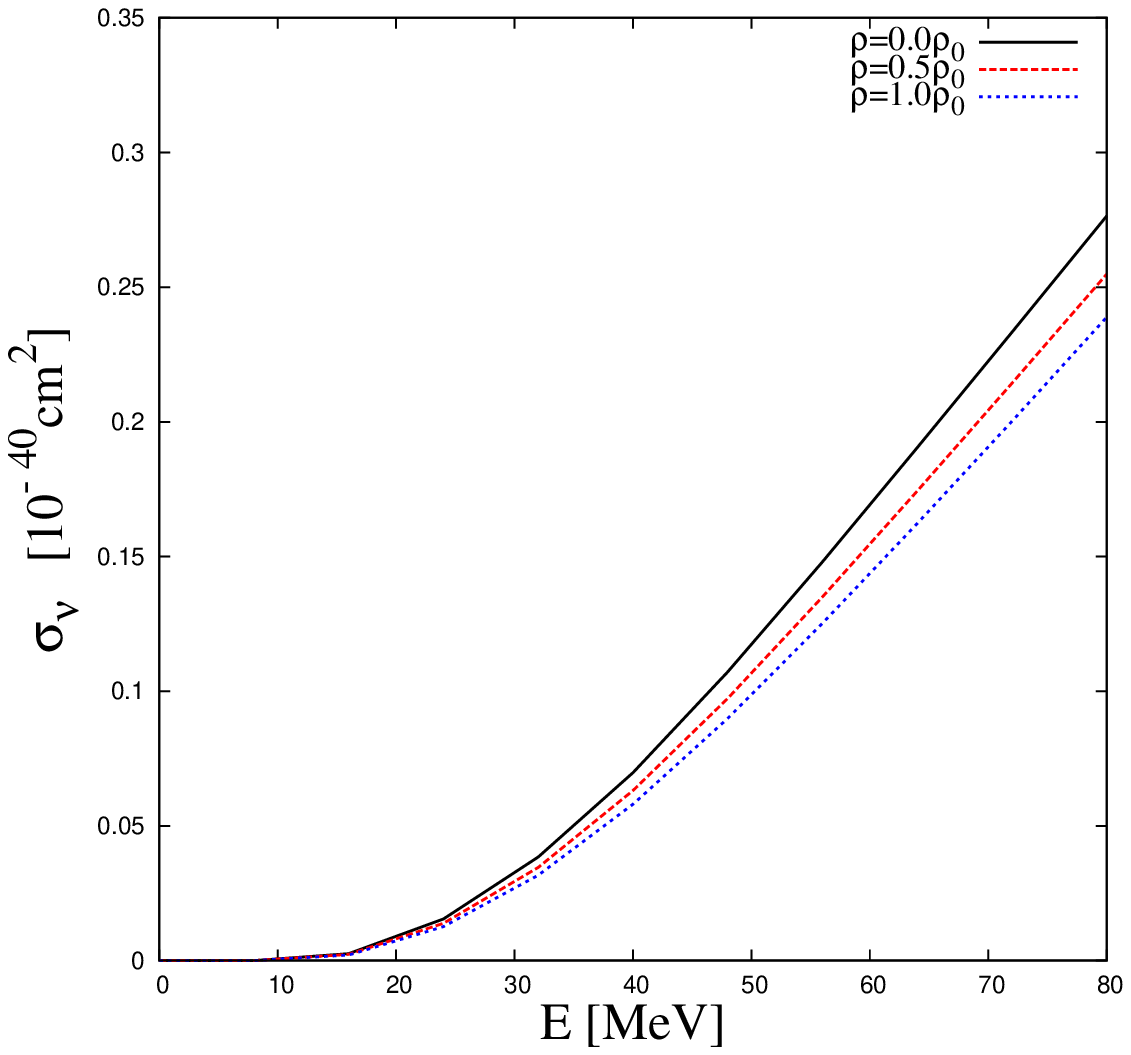}
\caption{(Color Online) Density dependence for the $^{12}$C$({\bar \nu}_e, {\bar \nu}_e^{'} ) ^{12}$C$^{*}(1^+)$ (left) and $^{12}$C$(\nu_e, \nu_e^{'} ) ^{12}$C$^{*} (1^+)$ (right) reactions. The y axis is $10^{-40} cm^2$, while the
x-axis is the incident neutrino energy in MeV. The cross sections decrease with increasing the nuclear density {\it i.e.} $\rho / \rho_o $ = 0.5 (red(dashed)) and  1.0 (blue(dotted)) in both reactions.} \label{fig4:c12NC}
\end{figure}
\begin{figure}
\centering
\includegraphics[width=8.0cm]{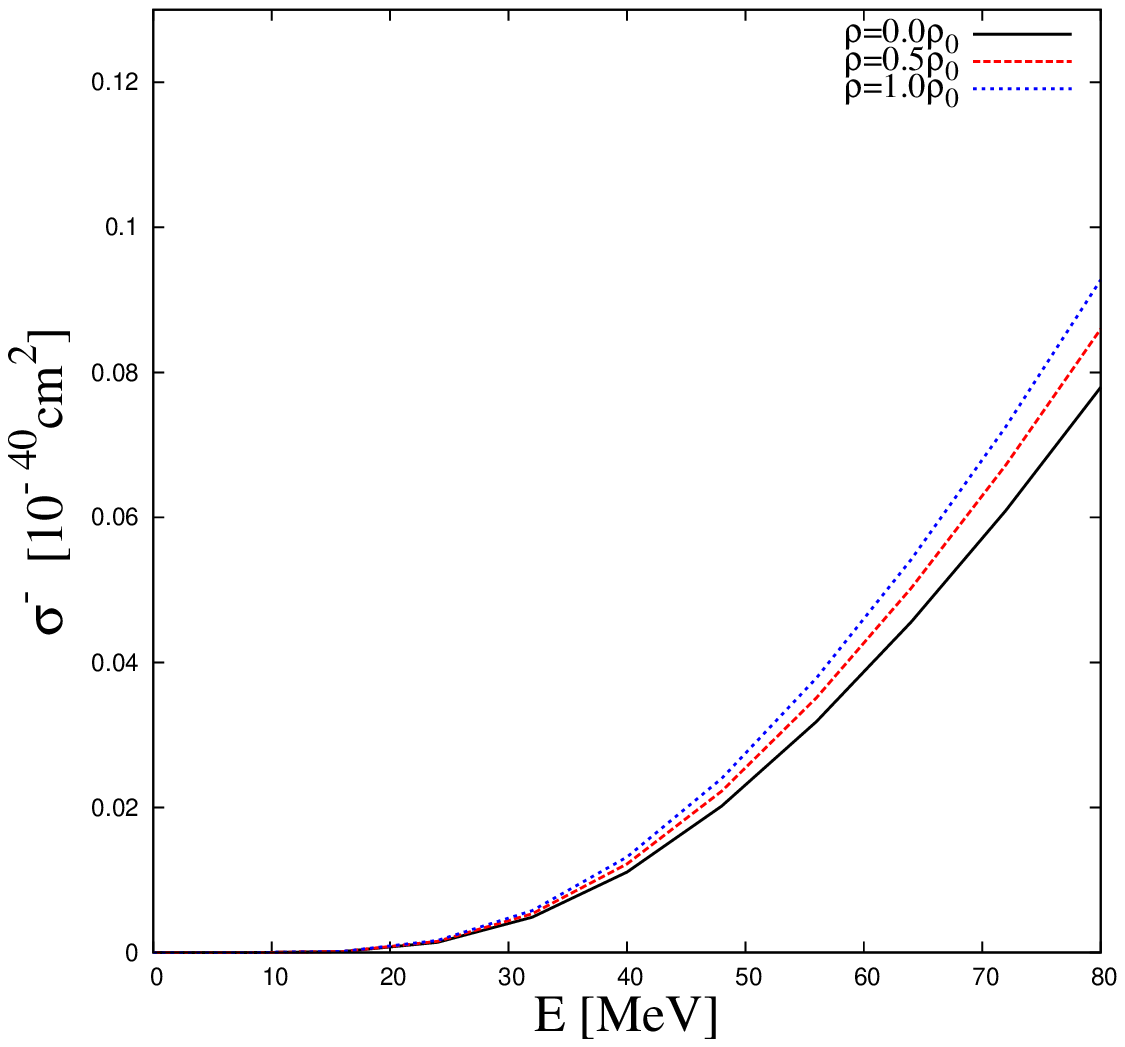}
\includegraphics[width=8.0cm]{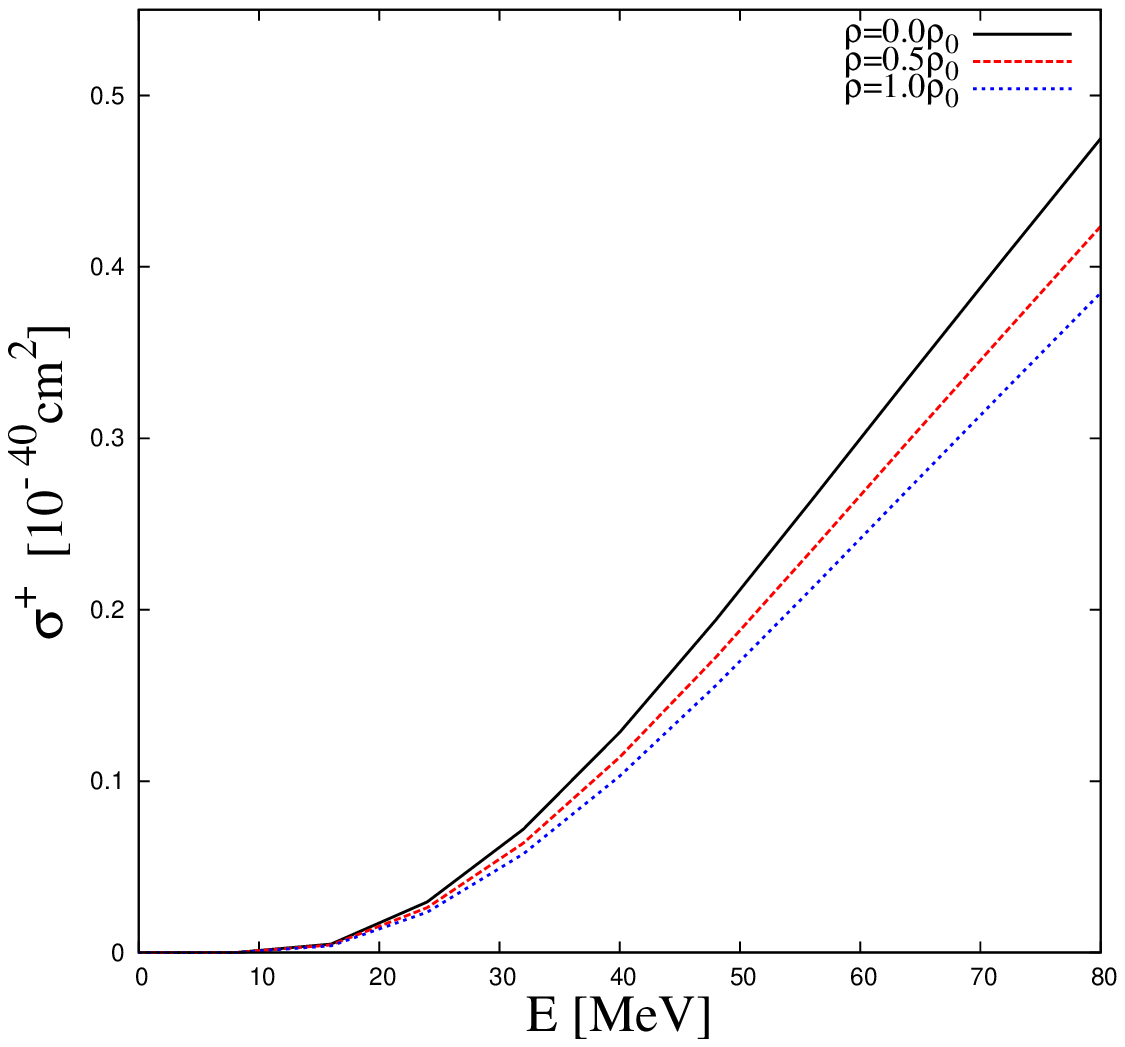}
\caption{(Color Online) Comparison of the total cross sections on $^{12}$C via the NC reaction
for $\sigma^- =\sigma (\nu_e) - \sigma ( {\bar \nu}_e)$ (left) and $\sigma^+ =\sigma (\nu_e) + \sigma ( {\bar \nu}_e)$ (right), which correspond to the HD and the HID parts in Eq.~(\ref{eq:DiffccC}).
Legends for the curves are the same as those for Fig.~\ref{fig4:c12NC}.} \label{fig5:asymC12NC}
\end{figure}

We have applied Eq.~(\ref{eq:DiffccC}) to the $^{12}$C$({\bar \nu}_e, {\bar \nu}_e^{'} ) ^{12}$C$^{*}(1^+)$ and $^{12}$C$(\nu_e, \nu_e^{'} ) ^{12}$C$^{*} (1^+)$ reactions. The reactions can be treated by the $\Delta J = 1$ transition from the $0^+$ ground state of $^{12}$C to the $1^+$ excited state. Descriptions of the nuclear states are performed by the QRPA framework \cite{ch10-2}. Our numerical results are presented in Fig.~\ref{fig4:c12NC}.
Medium effects on ${\bar \nu}_e$ and ${\nu}_e$ reactions on $^{12}$C appear
in a similar fashion to those on the nucleon in nuclear matter.

Total cross sections for the ${\nu_e}$ reaction decrease about 15\% per each density decrease of 0.5 $\rho_o$,
while those for the $\nu_e$ reaction are about 5\% for each decrease of the nuclear density.
If we take the average Fermi momentum of the nucleon in $^{12}$C, $k_F = 225 MeV (\rho = 0.668 \rho_o)$ \cite{Kim04},
the maximum change by the in-medium effects is shown to be less than 7\%.
Since we consider the exclusive reaction for the ground state of daughter nuclei, the cross sections are smaller than those for the nucleon in Fig.~\ref{fig2:nucleonNC}.

\begin{figure}
\centering
\includegraphics[width=8.0cm]{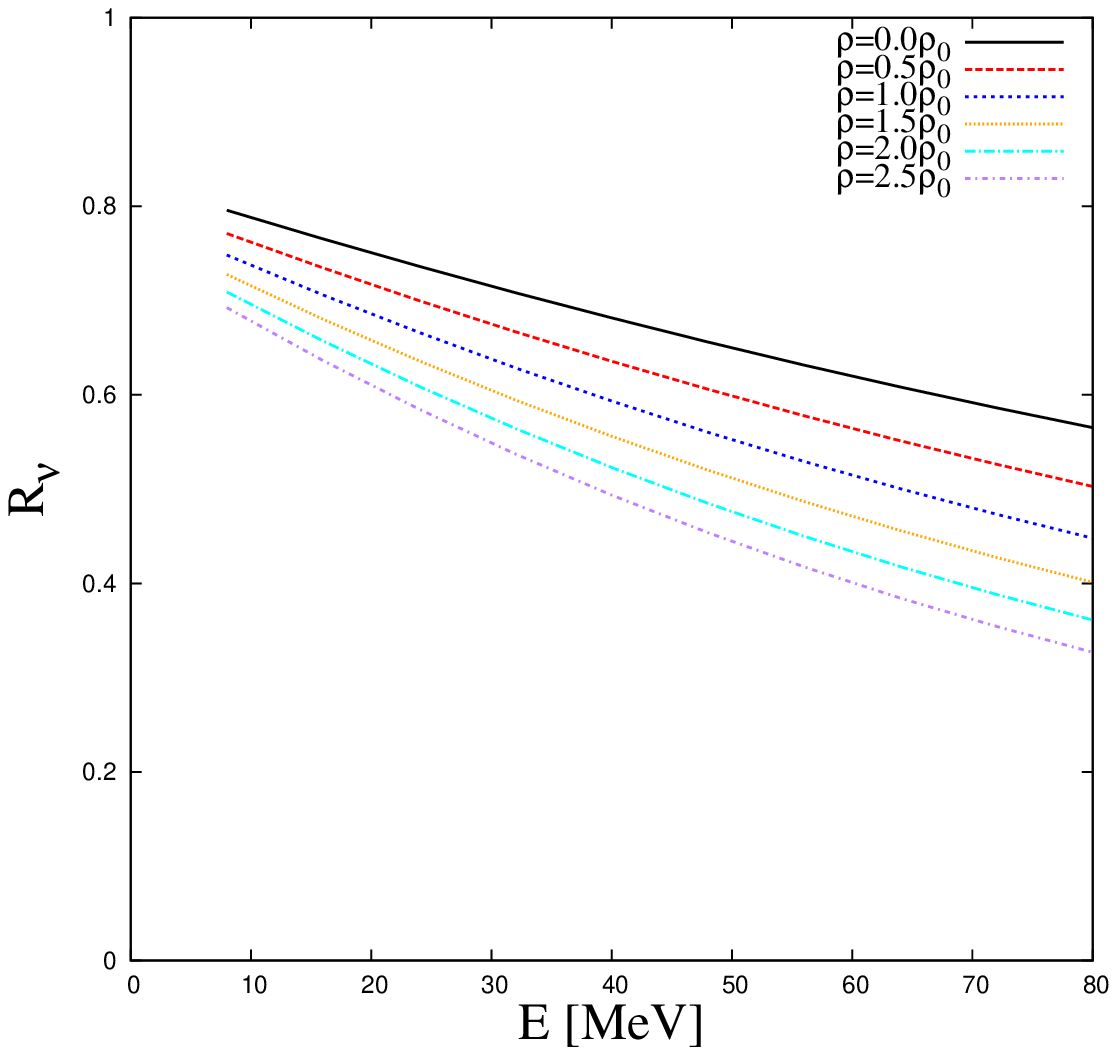}
\includegraphics[width=8.0cm]{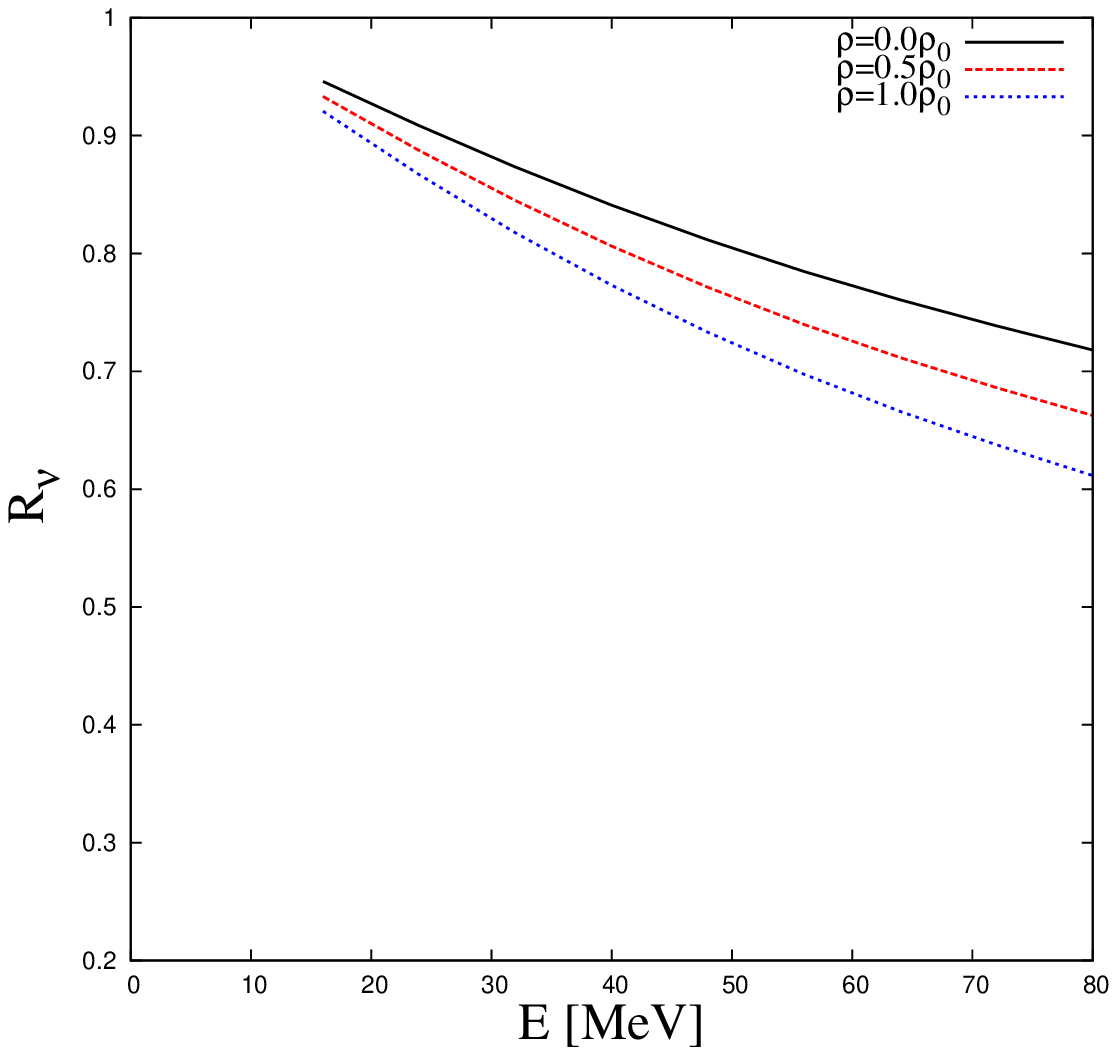}
\caption{(Color Online) Density and incident energy dependence of ${\bar \nu}_e$ and ${\nu}_e$ reaction (NC) ratio, $R_{\nu} = \sigma ({\bar \nu}_e) /  \sigma ({\nu}_e)$, on the nucleon (left) and $^{12}$C (right), which are the cross section ratios of left and right panels in Figs.\ref{fig2:nucleonNC} and \ref{fig4:c12NC}, respectively. Legends for the curves are same as Fig.\ref{fig4:c12NC}.} \label{fig6:ratioNC}
\end{figure}
In order to justify our approach to the NC neutrino reaction in finite nuclei, we compare our results to the experimental data, $10.4 \pm 1.0 \pm 0.9 \times 10^{-42} cm^2$, which was measured as the flux averaged cross section of $(\nu_e, \nu_e^{'}) + (\nu_{\mu}, \nu_{\mu}^{'})$ reactions on $^{12}$C target~\cite{Athan97}. If we take into account of the medium effect on $^{12}$C by adopting $\rho = 0.6 \rho_o$, our result become 9.52 $ \times 10^{-42} cm^2$~\cite{ch10}, which is consistent with the data. Comparison to the CC reactions data was done in Ref. \cite{DD1}.

In Fig.5, one may find that similar mechanism to the case of a nucleon in nuclear matter gives also rise to the asymmetry on the neutrino reaction on $^{12}$C in Fig.4. Namely the HD term, the last term in Eq.(10), calculated as $\sigma^-$ in the left panel of Fig.5 enhances (supresses) the medium effects on the ${\bar \nu}_e$ ($\nu_e$) reaction. Finally, in Fig.~\ref{fig6:ratioNC}, we show the cross section ratios, $\sigma ({\bar \nu}_e) /  \sigma ({\nu}_e)$,
on the nucleon in nuclear matter and $^{12}$C. All ratios decrease with increasing the incident energy $E_{\nu}$.
It would be a valuable low energy extension of the previous calculations performed
beyond $E_{\nu}$ = 100 MeV region~\cite{Kuzmin06}. As for the medium effects, the higher the density increases,
the smaller the ratios of the ${\bar \nu}_e$ to the $\nu_e$ reaction become. It means that the asymmetry between the ${\bar \nu}_e$ and ${\nu}_e$ reactions
would be increased in a denser nuclear medium. More careful treatment of the $\nu$ and ${\bar \nu}$ propagation in dense matter are necessary for more thorough understanding of the phenomena related to the neutrino propagation in nuclear matter.  

\section{Summary}

In summary, we applied the bound nucleon weak form factors modified in a nuclear medium to the neutrino and
the anti-neutrino reactions via neutral current on the nucleon in nuclear matter and $^{12}$C.
The form factors are calculated in the QMC model, and retain explicitly the four-momentum transfer and
the density dependence. Anti-neutrino reaction cross sections are largely suppressed in nuclear matter, i.e.,
about 30 \% maximal suppression around $\rho = \rho_0$ similarly to that observed for the CC reaction.
However, the neutrino cross sections in dense matter may be modified about 12 $\sim$ 18 \%, maximally.
Such asymmetry turns out to appear irrespective of the target. Therefore, it could affect significantly
the neutrino scattering, in particular, the anti-neutrino propagation inside the proto-neutron star.

Recent neutrino facilities present lots of fruitful data for the neutrino reaction in the quasi-elastic region \cite{Mini-CC-10,Mini-NC-10}. Although most of the data are now focused on the extraction of the axial mass and the strangeness content on a nucleon by the quasi-elastic kinematics, the study of the asymmetry between the neutrino and anti-neutrino reactions by more data on the ${\bar \nu}$ reaction could be the valuable alternative approach to understand the modification of the nucleon properties in a nuclear medium.




\acknowledgments

This work was supported by the National Research Foundation of Korea (Grant No. 2011-0003188).
The work of KT was supported by the University of Adelaide and the Australian Research Council through
grant No.~FL0992247 (Anthony William Thomas).

\newpage

\appendix

\section{Definitions of Form Factors}
In this paper, we use two different nucleon form factors. One is the Sachs form factors based on
the dipole form factor~\cite{musolf,Kim95}:
\begin{eqnarray}
G_D^V (Q^2)(& \equiv & {( 1 + Q^2 /M_V^2)}^{-2} ) =  (1 + 4.97 ~ \tau )^{-2} = G_E^{p,emp}  (Q^2)~,\\ \nonumber
~ G_E^{n,emp}  (Q^2) &=&-\mu_n \tau G_D^V  (Q^2) \eta ~ ,~ G_M^{p,emp} (Q^2)  =  \mu_p G_D^V  (Q^2)~,~G_M^{n,emp}  (Q^2)=\mu_n G_D^V (Q^2)~,
\end{eqnarray}
where $q = k_{\nu}^{i} - k_{l}^{f} = p_f - p_i, q^2 (= - Q^2) = q_0^2 ( =\omega^2) - {\bf q}^2 \le 0 $ with  $\tau =   Q^2/(4 {M_N^0}^2) = - q^2/(4 {M_N^0}^2) \ge 0, \eta = (1 +5.6 ~ \tau )^{-1}  $ and $M_N^0 = 0.939,\mu_p (= 1 + \lambda_p) =2.793, \mu_n (=\lambda_n) =-1.913$.

Note the following facts.

1) The $Q^2 = \omega^2 - {\bf q}^2$ defined in Eq.~(24) in Ref.~\cite{musolf} is correct,
but the argument $Q^2$ in the form factors is understood as $|Q^2|$ as shown in their figures.
In the same sense, the $G_D^V \equiv {( 1 - Q^2/M_V^2)}^{-2}$ in Eq.~(33c) is correct, {\it i.e.} $G_D^V \equiv {( 1 - |Q^2| /M_V^2)}^{-2}$. But it should be rewritten as $G_D^V \equiv {( 1 + Q^2/M_V^2)}^{-2}$ by our notation $Q^2 = {\bf q}^2 - \omega^2$.

2) Since $ Q^2/M_V^2 = 4.97 ~ \tau =  4.97 ~ Q^2/(4 {M_N^0}^2)$, $M_V^2$ is 0.71 GeV$^2$ ($M_V$ = 0.84 GeV), which is consistent with the standard value~\cite{Weise}.


The Sachs form factors are related to the Dirac and Pauli form factors as follows
\begin{eqnarray}
F_1^{p,emp} (Q^2) &=& (G_E^{p,emp}  (Q^2)+ \tau G_M^{p,emp}  (Q^2) )/(1+\tau) = [ 1 + \tau ( 1 + \lambda_p)] G_D^V (Q^2) / ( 1 + \tau)~, \nonumber \\ 
F_1^{n,emp} (Q^2) &=& (G_E^{n,emp}  (Q^2)+ \tau G_M^{n,emp}  (Q^2) )/(1+\tau) =  \tau \lambda_n ( 1  - \eta) G_D^V (Q^2) / ( 1 + \tau)~,  \nonumber \\
F_2^{p,emp} (Q^2) &= & (G_M^{p,emp}  (Q^2)-G_E^{p,emp}  (Q^2))/(1+ \tau) = \lambda_p G_D^V (Q^2) / ( 1 + \tau) ~,  \nonumber \\
F_2^{n,emp} (Q^2) &= & (G_M^{n,emp}  (Q^2)-G_E^{n,emp}  (Q^2))/(1+ \tau) = \lambda_n ( 1 + \tau \eta) G_D^V (Q^2) / ( 1 + \tau)~, 
\end{eqnarray}
vice verse
\begin{equation}
G_E^{p(n),emp} (Q^2) = F_1^{p(n),emp} (Q^2) - \tau F_2^{p(n),emp}  (Q^2)~,~G_M^{p(n),emp}  (Q^2)=F_1^{p(n),emp} (Q^2) + F_2^{p(n),emp}  (Q^2).
\end{equation}
The isovector-vector weak form factors are given by the Dirac and Pauli form factors,
\begin{eqnarray}
F_{1,2}^{V}  (Q^2)&=&F_{1,2}^{p,emp}  (Q^2)-F_{1,2}^{n,emp}  (Q^2)~ (CC)~,~ \\ \nonumber
F_{1,2}^{V,p(n)}  (Q^2)&=&( {1 \over 2} - 2 sin^2 \theta_w) F_{1,2}^{p(n)}  (Q^2)- { 1 \over 2} F_{1,2}^{n(p)}  (Q^2)~ (NC)~,~
\end{eqnarray}
where the NC case is taken from our previous paper~\cite{ch10} and Ref.~\cite{giusti1}.
On the other hand, the axial vector form factor is given by $G_A^{emp}=g_A^{cc} \times (1 + Q^2/M_A^2)^{-2}$
with $M_A=1.03$ GeV and $g_A=1.26$.

Other form factors introduced in Refs.~\cite{Donn72,Paar08,Wale75} are all assumed to have the same momentum dependence
in MeV units:
\begin{eqnarray}
F_{1}^V (q^2) &=& {(1 + q^2 / {(855 MeV)}^2)}^{-2}~,~ F_A (q^2)= -1.23 \times {( 1 + q^2 / {(855 MeV)}^2)}^{-2} ~,\\ \nonumber
\mu_V (q^2) &=& F_1^V (q^2)+ 2 M_N F_2^V (q^2)= 4.706 \times {(1 + q^2 /{( 855 MeV)}^2)}^{-2}~,~ \\ \nonumber
F_P(q^2) &=& 2 M_N F_A (q^2) /( q^2 + m_{\pi}^2)~,
\end{eqnarray}
with $m_{\pi}$ = 139.57 MeV, $M_N$ = 931.49432 MeV, $\mu_V (0) = \mu_p - \mu_n = 4.706$.

Note the following facts.

1) Here $q^2 = {\bf q}^2 - q_0^2$ defined by Eq.~(80) in Ref.~\cite{Wale75} is the same $Q^2$ definition with our notation.
The $F_1^V ( q^2)$ in A(5) is almost same as that in (A4) because $F_1^V (q^2) = F_1^{p}  (Q^2)-F_1^{n} (Q^2) \simeq G_D^V(q^2) = {(1 + Q^2 / {(855 MeV)}^2)}^{-2} $ in the
low momentum transfer region. But, $F_A (q^2)$ is a bit different from the standard one, {\it i.e.}
$M_A = 0.855$ GeV and $g_A$ = --1.23 in (A5) is smaller than $M_A = 1.03$ GeV and $g_A$ = 1.26, where the ``minus'' sign comes from the different sign in the axial part of the weak current.

2) If we define $2M_N F_2^V (q^2) = F_2^V (q^2) $ which depends on the definition of the vector current, $\mu_V (q^2)$ in (A5) is equal to $G_M^V (q^2)$ in Eq.(5) because $\mu_V( q^2) = F_1^V (q^2) + F_2^V (q^2) \simeq (1 + \lambda_p - \lambda_n) G_D^V ( q^2) = (\mu_p - \mu_n) G_D^V ( q^2) = 4.706 \times {(1 + q^2 /{( 855 MeV)}^2)}^{-2} $ in the low momentum transfer region.

\section{Density dependent Form Factors}

Here, we show the density dependence of the various form factors calculated in the QMC model.
In Figs.~7-10, $G_E^p (\rho, Q^2),G_E^n (\rho, Q^2),G_M^p (\rho, Q^2)$ and $G_M^n (\rho, Q^2)$ form factors and their ratios $R(G_{E,M}^{p,n}) = G_{E,M}^{p,n} (\rho, Q^2) / G_{E,M}^{p,n} (\rho = 0, Q^2)$ are presented.
Ratios of the electric form factors in Figs.~7 and 8 converge to 1.0 at $Q^2$ = 0,
but those of the magnetic form factors in Figs.~9 and 10 are enhanced. The axial vector form factor is quenched in a nuclear medium, even at $Q^2$ = 0, which causes the change of the neutrino reaction in dense matter
in the cosmos, albeit their small momentum transfer. Ratios of density-dependent weak form factors, $F_{1,2}^V$, are presented in Fig.~12. About a 25 \% increase of $F_2^V$ is to be noticed for the weak interaction
in dense matter.

But in the neutrino reaction on the quasi-elastic region, for example, the kinematical region performed
at MiniBooNE experiments \cite{Mini-CC-10,Mini-NC-10}, the dependence on the 4-momentum transfer $Q^2$, as well as that on the
nuclear density, may be practically important.

\newpage

\begin{figure}
\centering
\includegraphics[width=8.0cm]{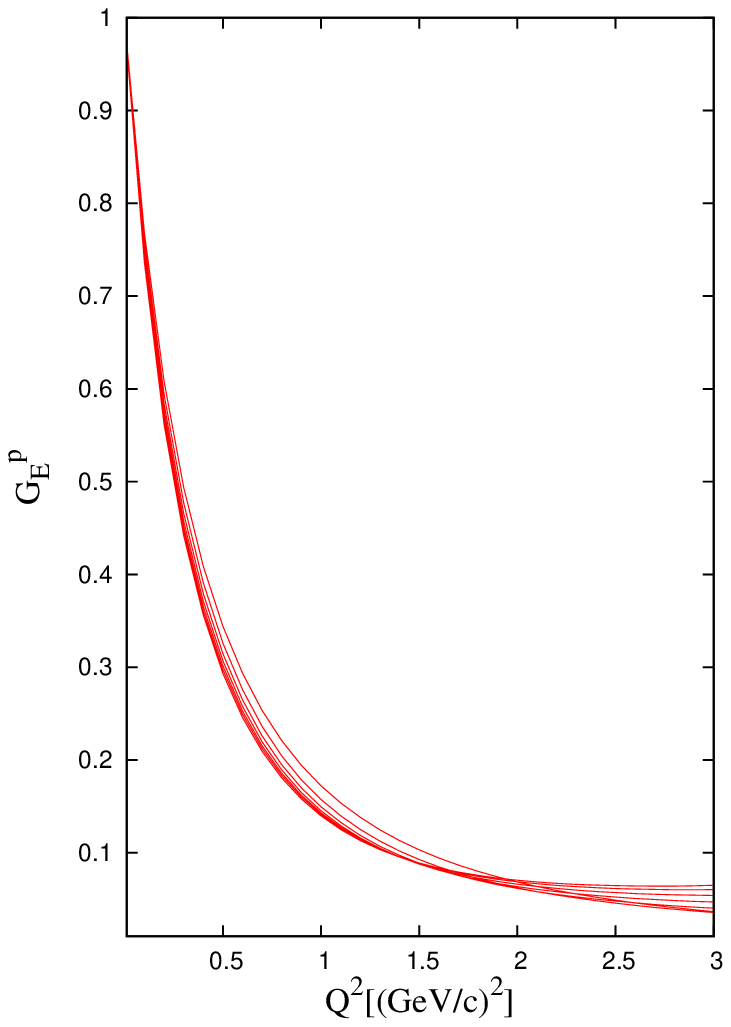}
\includegraphics[width=8.0cm]{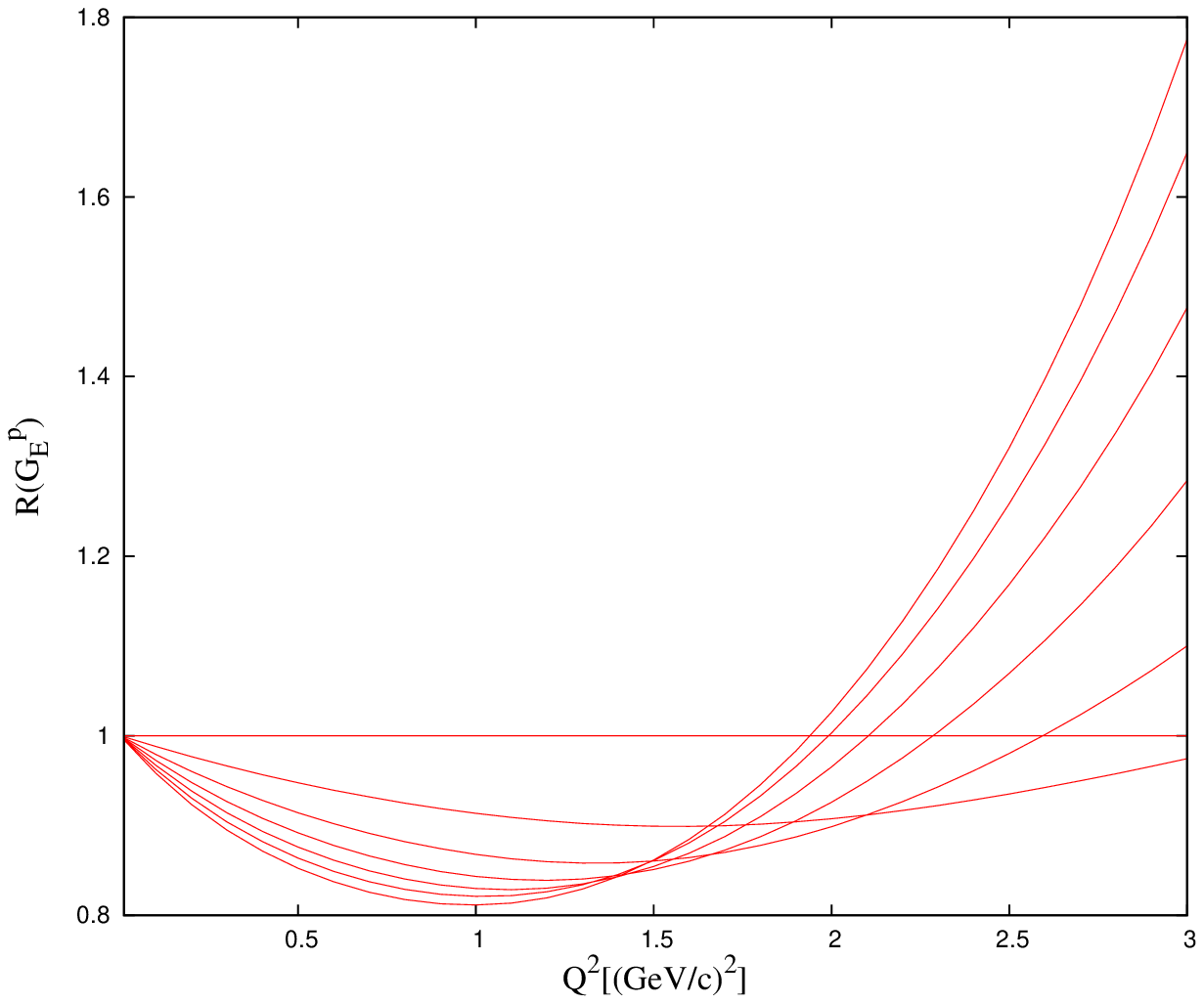}
\caption{(Color online) $G_E^p (\rho, Q^2)$ and $R (G_E^p) = G_E^p (\rho, Q^2) / G_E^p (\rho = 0, Q^2) $ in nuclear matter.
The uppermost curves at $Q^2$ = 1.5 ${[GeV/c]}^2$ are for $\rho$ = 0, from which density increases by 0.5 $\rho_o$. } \label{figA1:GEP}
\end{figure}
\begin{figure}
\centering
\includegraphics[width=8.0cm]{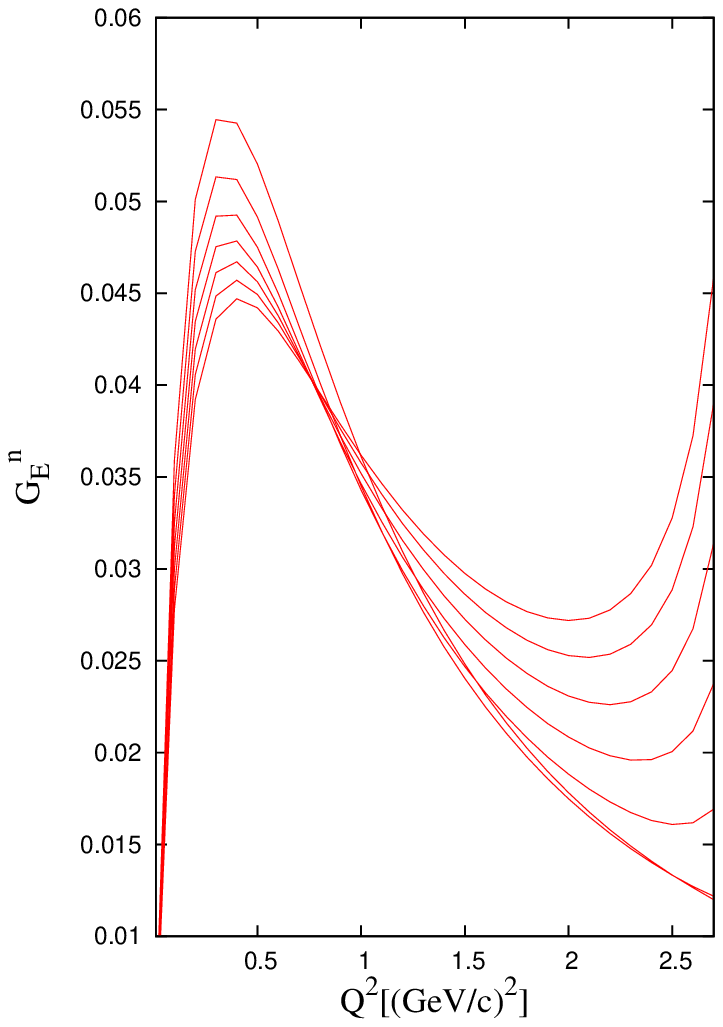}
\includegraphics[width=8.0cm]{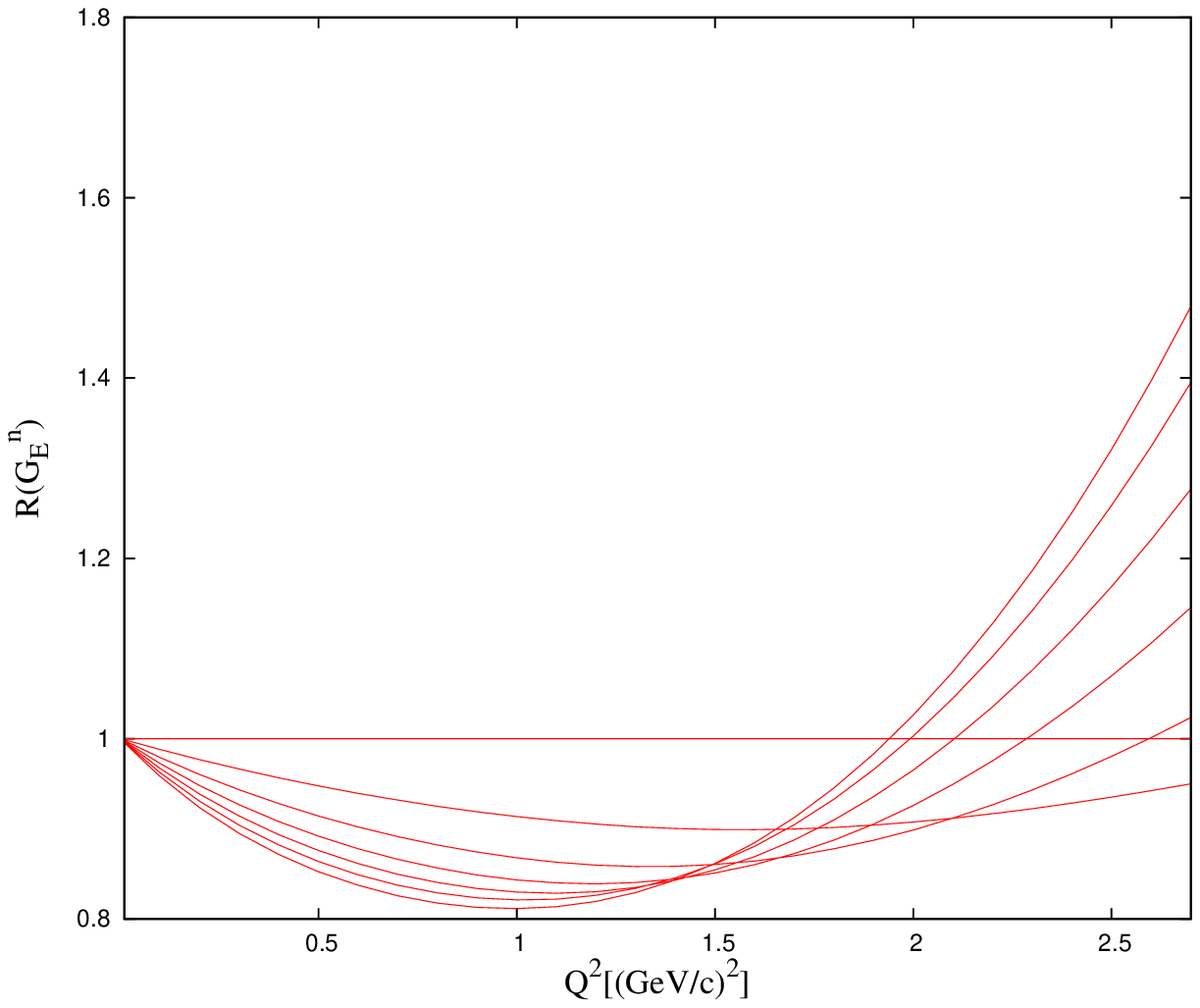}
\caption{(Color online) $G_E^n (\rho, Q^2)$ and $R (G_E^n) = G_E^n (\rho, Q^2) / G_E^n (\rho = 0, Q^2) $ in nuclear matter.
The uppermost curves at $Q^2$ = 0.5 ${[GeV/c]}^2$ are for $\rho$ = 0, from which density increases by 0.5 $\rho_o$. }
\label{figA1:GEN}
\end{figure}
\begin{figure}
\centering
\includegraphics[width=8.0cm]{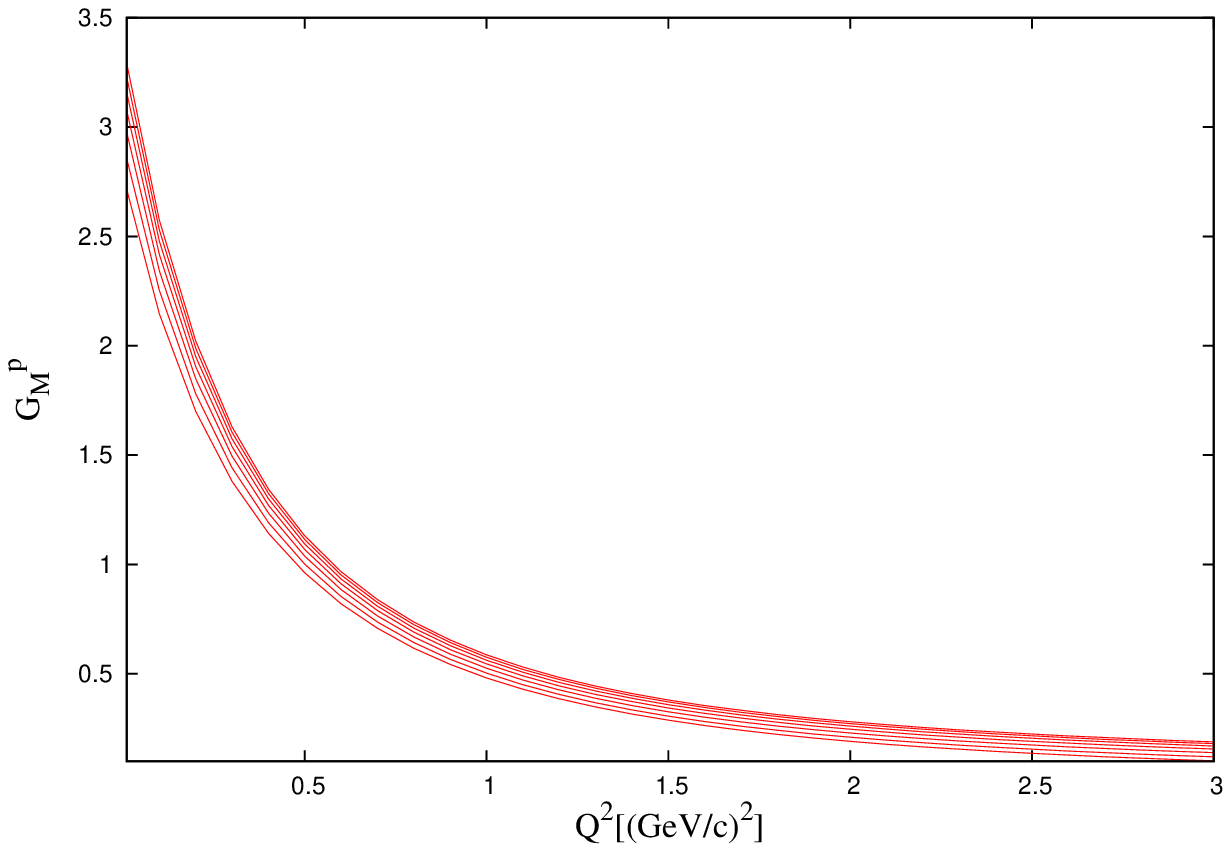}
\includegraphics[width=8.0cm]{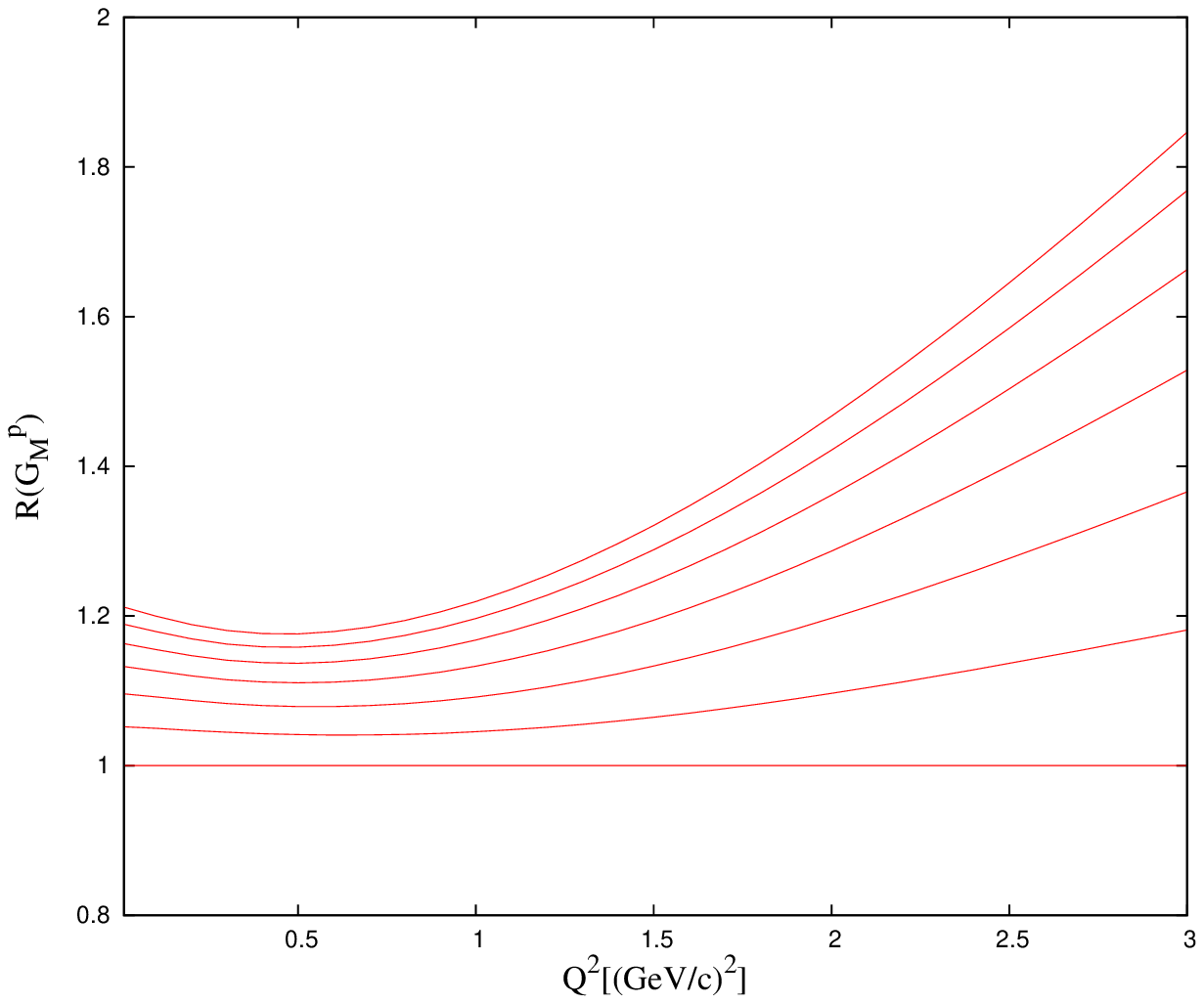}
\caption{(Color online) $G_M^p (\rho, Q^2)$ and $R (G_M^p) = G_M^p (\rho, Q^2) / G_M^p (\rho = 0, Q^2) $ in nuclear matter. The lowermost curves are for $\rho$ = 0, from which density increases by 0.5 $\rho_o$. } \label{figA1:GMP}
\end{figure}
\begin{figure}
\centering
\includegraphics[width=8.0cm]{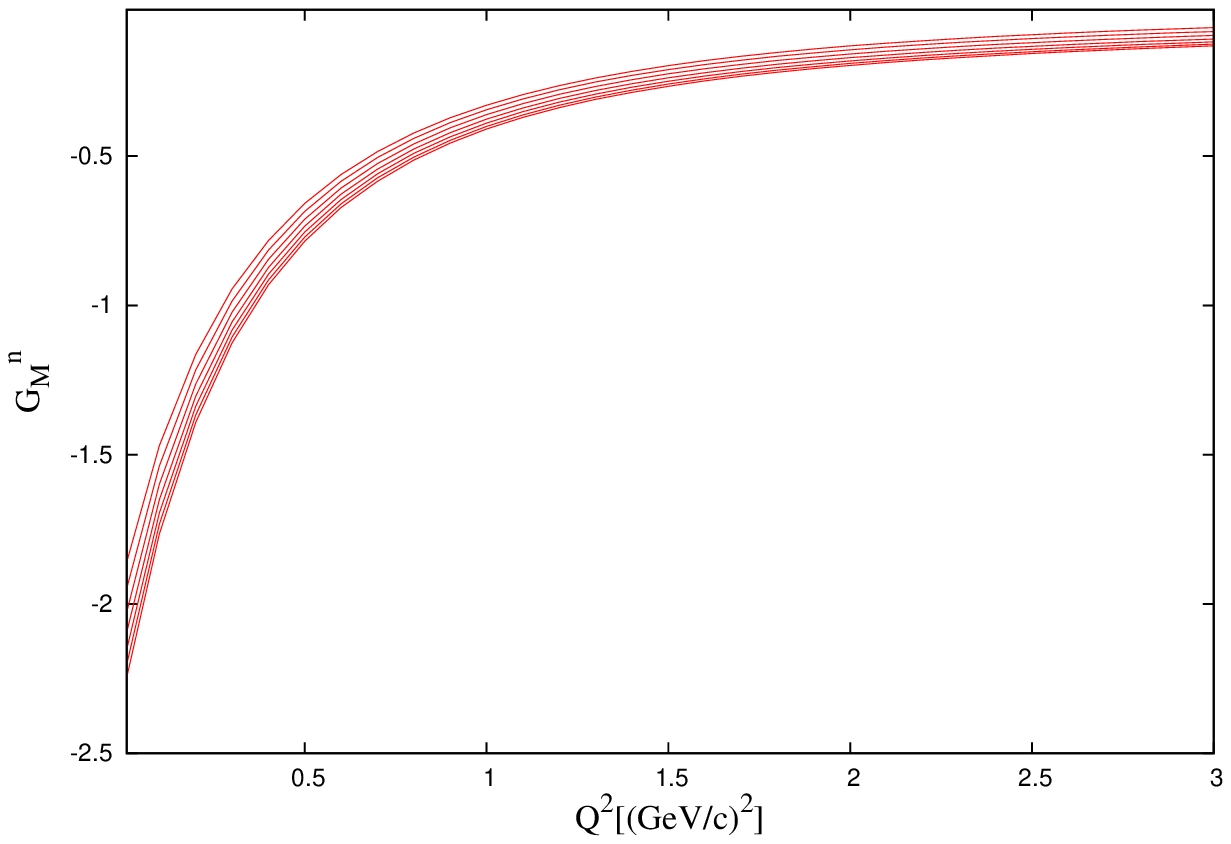}
\includegraphics[width=8.0cm]{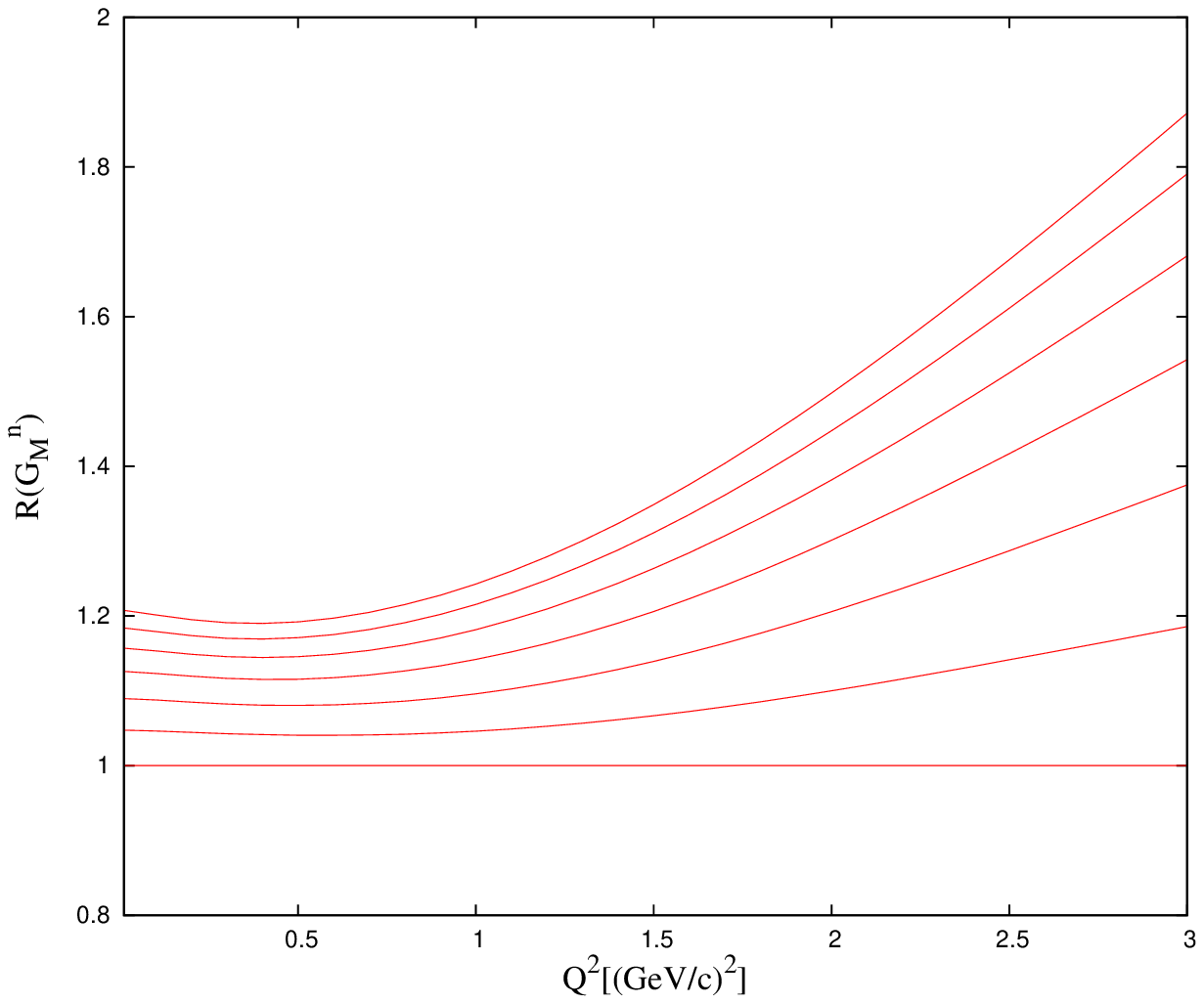}
\caption{(Color online) $G_M^n (\rho, Q^2)$ and $R (G_M^n) = G_M^n (\rho, Q^2) / G_M^n (\rho = 0, Q^2) $ in nuclear matter. The uppermost (lowermost) curve in left (right) is for $\rho$ = 0, from which density is increases by 0.5 $\rho_o$. } \label{figA1:GMN}
\end{figure}
\begin{figure}
\centering
\includegraphics[width=8.0cm]{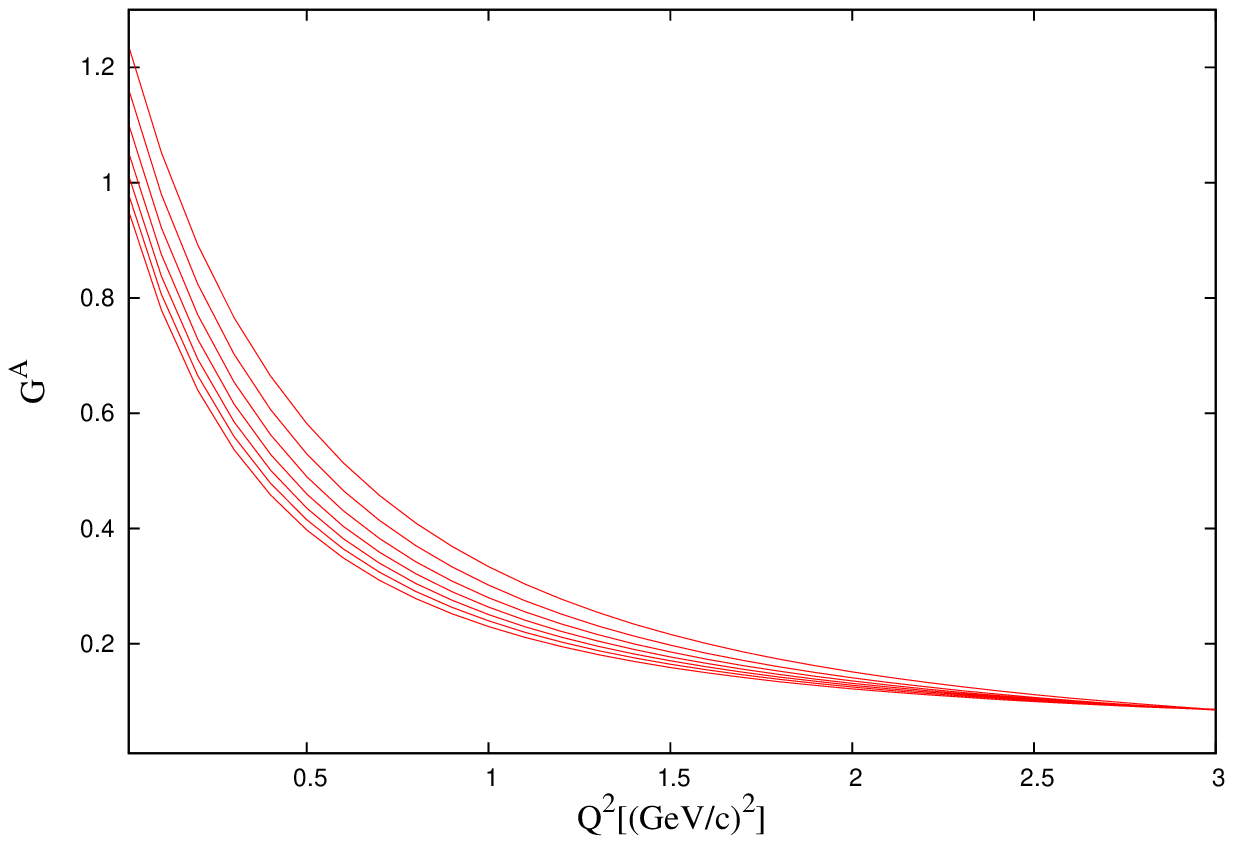}
\includegraphics[width=8.0cm]{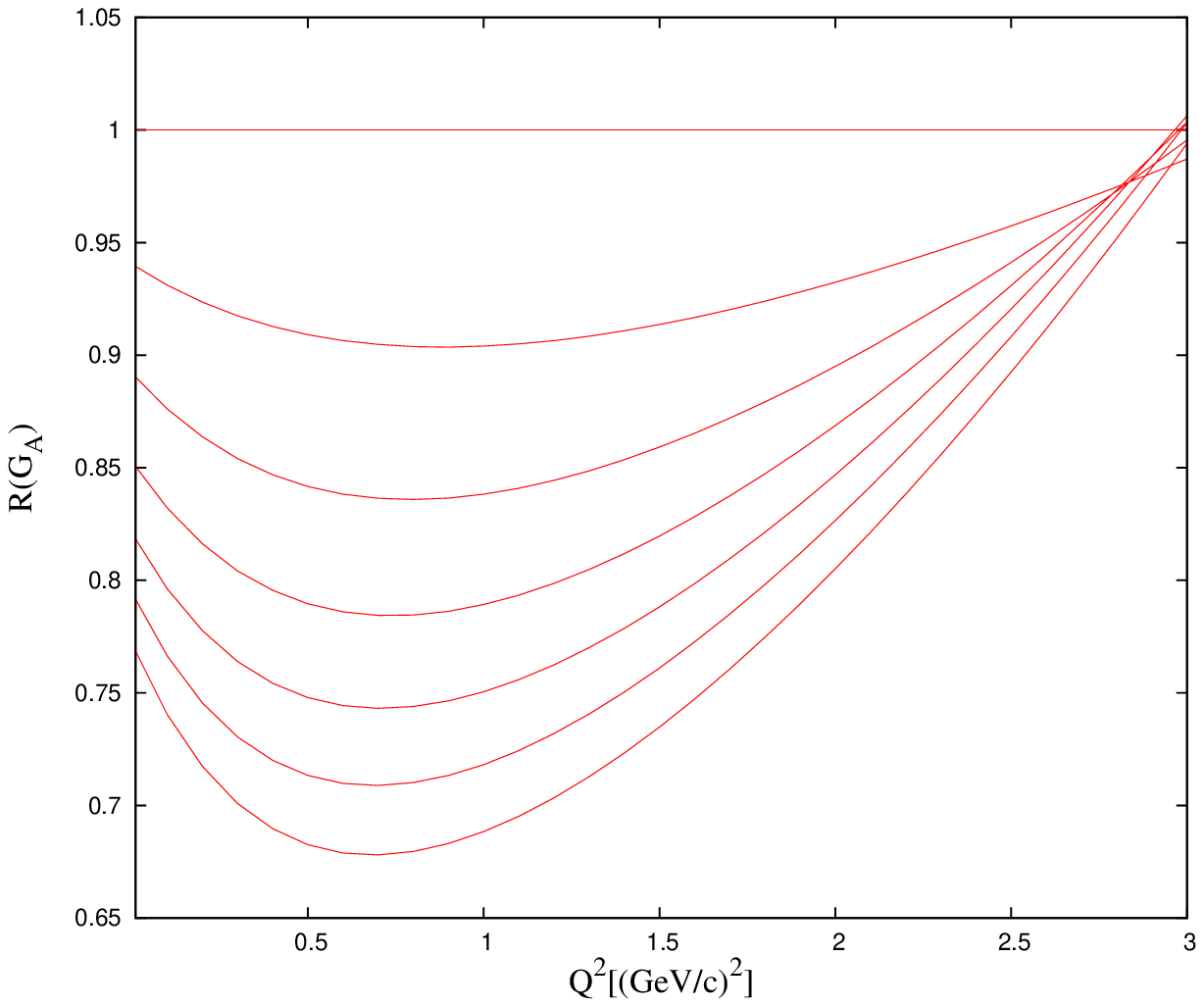}
\caption{(Color online) $G_A (\rho, Q^2)$ and $R (G_A) = G_A (\rho, Q^2) / G_A (\rho = 0, Q^2) $ in nuclear matter.
The uppermost curves are for $\rho$ = 0, from which density increases by 0.5 $\rho_o$. } \label{figA1:GA}
\end{figure}

\begin{figure}
\centering
\includegraphics[width=8.0cm]{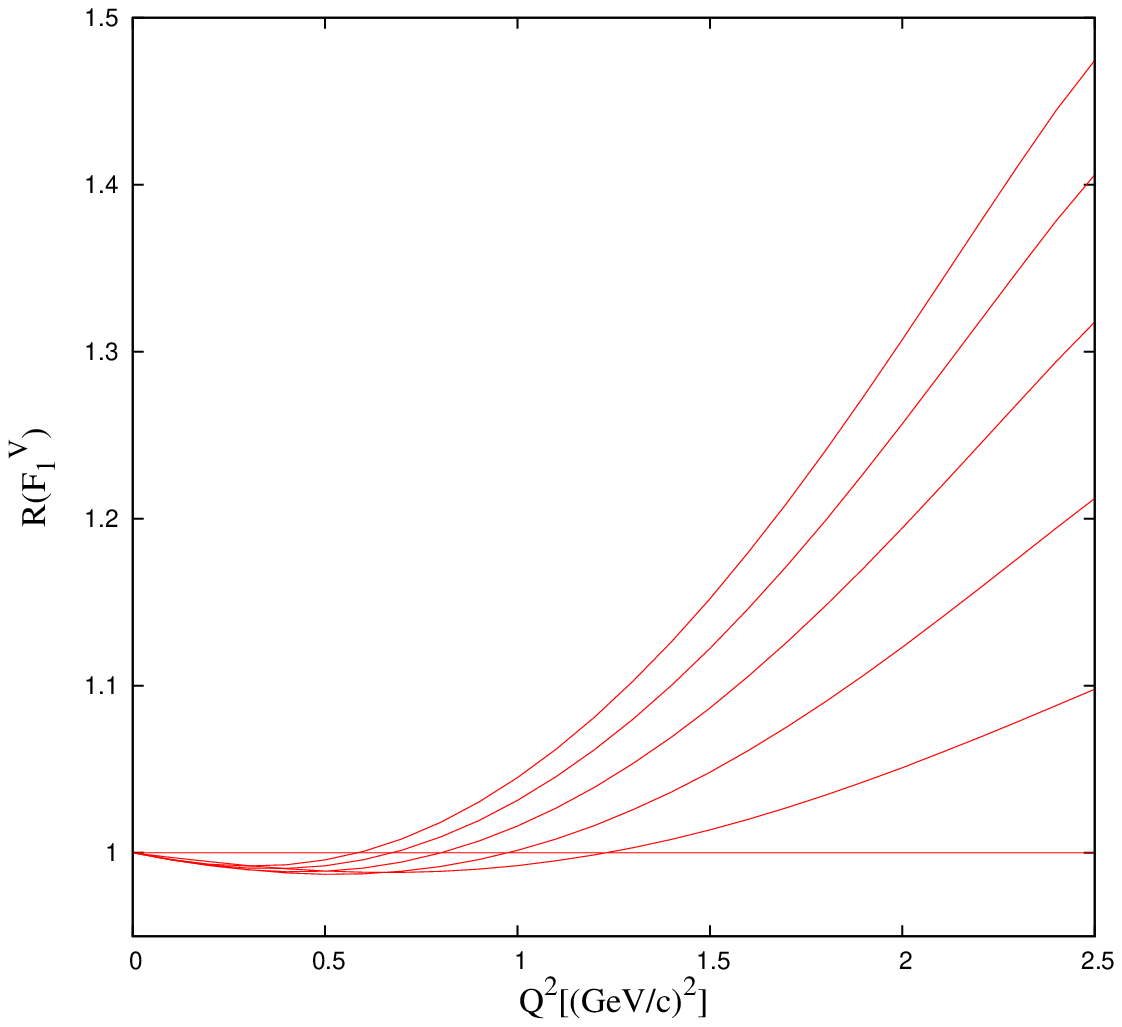}
\includegraphics[width=8.0cm]{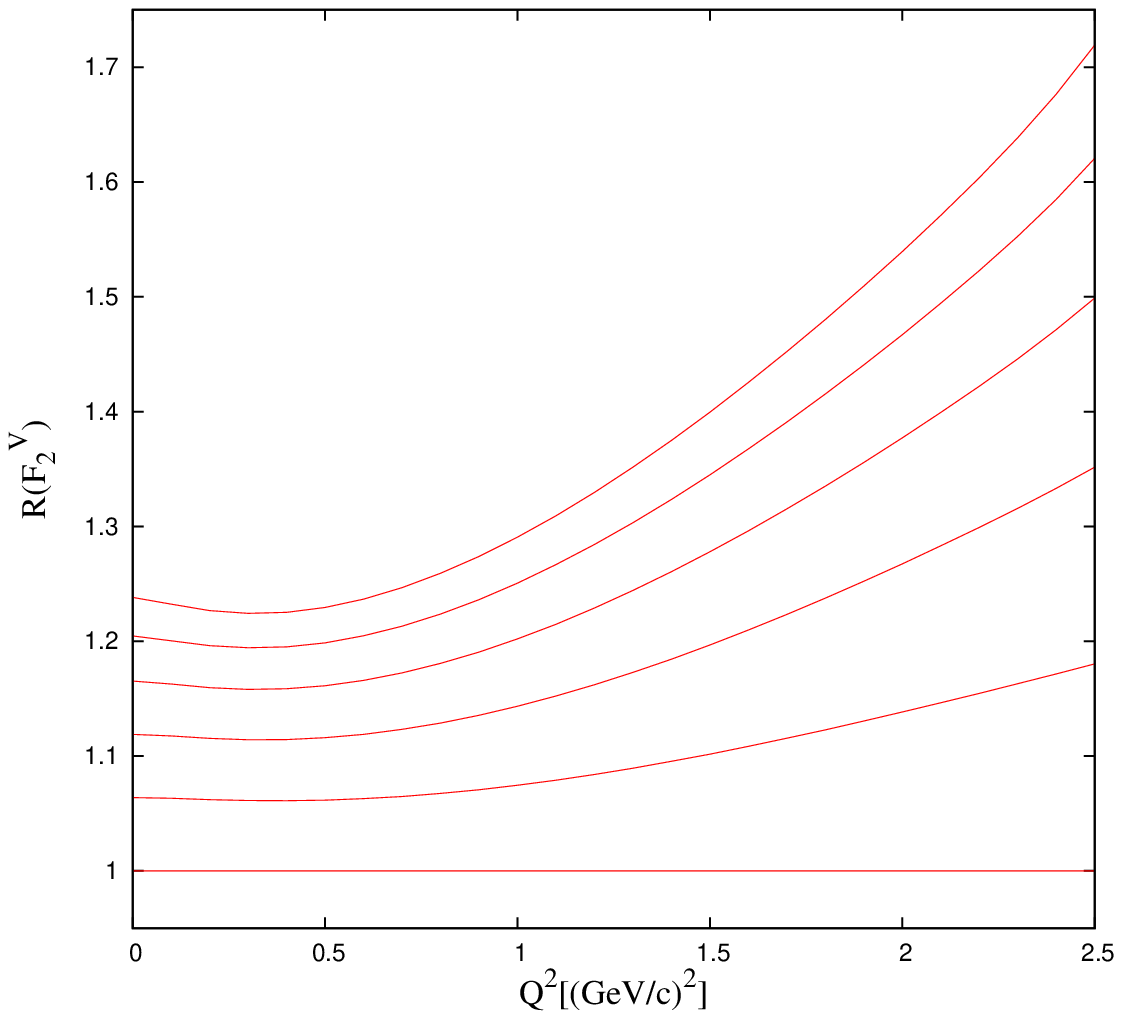}
\caption{Modification in the weak vector form factors, $ R(F_{1,2}^V) = F_{1,2}^{V} (\rho, Q^2)/F_{1,2}^{V}(\rho = 0, Q^2)$,
with finite momentum transfer in nuclear matter. From the lowermost (vacuum), density ratios increase by 0.5 $\rho_o$.
The uppermost curve is for $\rho = 2.5 \rho_o$. } \label{fig2:f1f2}
\end{figure}


\thebibliography{99}
\bibitem{Fisc11} T. Fischer, I. Sagert, G. Pagliara, M. Hempel, J. Schaffner-Bielich, T. Rauscher, F.-K. Thielemann, R. Kaeppeli, G. Martinez-Pinedo, and M. Liebendoerfer, Astrophys. J. Supplement {\bf 194}, 39 (2011).
\bibitem{Maru11} Tomoyuki Maruyama, Toshitaka Kajino, Nobutoshi Yasutake, Myung-Ki Cheoun, Chung-Yeol Ryu, Phys. Rev. {\bf D 83}, 081303 (2011).
\bibitem{Maru12} Tomoyuki Maruyama, Nobutoshi Yasutake, Myung-Ki Cheoun, Jun Hidaka, Toshitaka Kajino, Grant J. Mathews, Chung-Yeol Ryu, Phys. Rev. {\bf D 86}, 123003 (2012).
\bibitem{Maru13} Tomoyuki Maruyama, Jun Hidaka, T. Kajino, Nobutoshi Yasutake, T. Kuroda, Myung-Ki Cheoun, Chung-Yeol Ryu, Grant J. Mathews, arXiv:1301.7495[astro-ph.SR] (2013).
\bibitem{Wana11} Shinya Wanajo, Hans-Thomas Janka, Shigeru Kubono,
Astrophys. J. {\bf 729}, 46 (2011).



\bibitem{Ch10-2} Myung-Ki Cheoun, Eunja Ha, T. Hayakawa, Toshitaka Kajino, Satoshi Chiba, Phys. Rev. {\bf C82}, 035504 (2010).

\bibitem{Ch12} Myung-Ki Cheoun, Eunja Ha, T. Hayakawa, Satoshi Chiba, Ko Nakamura, Toshitaka Kajino, Grant J. Mathews, Phys. Rev. {\bf C85}, 065807 (2012).

\bibitem{Yosh08} T. Yoshida, T. Suzuki, S, Chiba, T. Kajino, H. Yokomukura,
K. Kimura, A. Takamura, H. Hartmann, Astrophys. J. {\bf 686}, 448 (2008).

\bibitem{Malo00} S. Malov, K. Wijesooriya, F. T. Baker, L. Bimbot,
E. J. Brash, C. C. Chang, J. M. Finn, K. G. Fissum
et al., Phys. Rev. {\bf C 62}, 057302 (2000).
\bibitem{Diet01} S. Dieterich, P. Bartsch, D. Baumann, J. Bermuth,
K. Bohinc, R. Bohm, D. Bosnar, S. Derber et al., Phys.
Lett. {\bf B 500}, 47-52 (2001).
\bibitem{Stra03} S. Strauch et al. [ Jefferson Lab E93-049 Collaboration ],
Phys. Rev. Lett. {\bf 91}, 052301 (2003).
\bibitem{Paol10} M. Paolone, S. P. Malace, S. Strauch, I. Albayrak, J. Arrington,
B. L. Berman, E. J. Brash, B. Briscoe et al.,
Phys. Rev. Lett. {\bf 105}, 072001 (2010).
\bibitem{Mala11} S. P. Malace, M. Paolone, S. Strauch, I. Albayrak, J. Arrington,
B. L. Berman, E. J. Brash, B. Briscoe et al.,
Phys. Rev. Lett. {\bf 106}, 052501 (2011).
%
\bibitem{Brooks:2011sa}
  For a review, W.~K.~Brooks, S.~Strauch and K.~Tsushima,
  J.\ Phys.\ Conf.\ Ser.\  {\bf 299}, 012011 (2011).
%
\bibitem{Cloe09} I. C. Cloet, G. A. Miller, E. Piasetzky and G. Ron, Phys.
Rev. Lett. {\bf 103}, 082301 (2009).

\bibitem{DD1} Myung-Ki Cheoun, Ki-Seok Choi, K.S. Kim, Koichi Saito, Toshitaka Kajino, Kazuo Tsushima and Tomoyuki Maruyama, arXiv:1302.5770 [nucl-th], (2013).



\bibitem{QMCboundff}
  D.~-H.~Lu, K.~Tsushima, A.~W.~Thomas, A.~G.~Williams and K.~Saito,
  Phys.\ Rev.\ C {\bf 60}, 068201 (1999).

\bibitem{QMChe3ff}
  D.~-H.~Lu, K.~Tsushima, A.~W.~Thomas, A.~G.~Williams and K.~Saito,
  Phys.\ Lett.\ B {\bf 441}, 27 (1998).

\bibitem{QMCmatterff}
  D.~-H.~Lu, A.~W.~Thomas, K.~Tsushima, A.~G.~Williams and K.~Saito,
  Phys.\ Lett.\ B {\bf 417}, 217 (1998).

\bibitem{QMCmatterga}
  D.~-H.~Lu, A.~W.~Thomas and K.~Tsushima,
  nucl-th/0112001 (2011).


\bibitem{QMCfinite}
K.~Saito, K.~Tsushima and A.~W.~Thomas,
  Nucl.\ Phys.\ A {\bf 609}, 339 (1996):\\
K.~Saito, K.~Tsushima and A.~W.~Thomas,
  Phys.\ Rev.\ C {\bf 55}, 2637 (1997).


\bibitem{QMChyp}
K.~Tsushima, K.~Saito, J.~Haidenbauer and A.~W.~Thomas,
  Nucl.\ Phys.\ A {\bf 630}, 691 (1998):\\
P.~A.~M.~Guichon, A.~W.~Thomas and K.~Tsushima,
  Nucl.\ Phys.\ A {\bf 814}, 66 (2008):\\
K.~Tsushima and F.~C.~Khanna,
  Phys.\ Rev.\ C {\bf 67}, 015211 (2003):\\
K.~Tsushima and F.~C.~Khanna,
  J.\ Phys.\ G {\bf 30}, 1765 (2004).


\bibitem{QMCreview}
  K.~Saito, K.~Tsushima and A.~W.~Thomas,
  Prog.\ Part.\ Nucl.\ Phys.\  {\bf 58}, 1 (2007).  

\bibitem{Kim04} K. Tsushima, Hungchong Kim, and K. Saito, Phys. Rev. {\bf C 70}, 038501 (2004).





\bibitem{ch10} Myung-Ki Cheoun, Eunja Ha, K.S. Kim, Toshitaka Kajino, J. Phys. {\bf G 37}, 055101 (2010).
\bibitem{ch10-2} Myung-Ki Cheoun, Eunja Ha, Su Youn Lee, K.S. Kim, W.Y. So, Toshitaka Kajino, Phys. Rev. {\bf C81}, 028501 (2010).
\bibitem{Ch-08} Myung-Ki Cheoun and K.S. Kim,  J. Phys. {\bf G 35}, 065107 (2008).
\bibitem{giusti1} Andrea Meucci, Carlotta Giusti, and Franco Davide
Pacati, Nucl. Phys. {\bf A739}, 277 (2004); Nucl. Phys. {\bf
A744}, 307 (2004); Nucl. Phys. {\bf A773}, 250 (2006).
\bibitem{musolf}M. J. Musolf and T. W. Donnelly, Nucl. Phys. {\bf
A546}, 509 (1992).
\bibitem{Kim95} Hungchong Kim, J. Piekarewicz and C. J. Horowitz, Phys. Rev. {\bf C} 51, 2739(1995).


\bibitem{Ch08} Myung-Ki Cheoun and K.S. Kim, J. Phys. {\bf G 35}, 065107 (2008).

\bibitem{Albe02} W. M. Alberico, S. M. Bilenky, C. Maieron, Phys.
Rep. {\bf 358}, 227-308, (2002).

\bibitem{Stru03} Alessandro Strumia and Francesco Vissani, Phys. Lett. {\bf B 564}, 42 (2003).


\bibitem{Athan97} C. Athanassopoulos et al. (LSND Collaboration), Phys. Rev. {\bf C
55}, 2078 (1997).


\bibitem{Kuzmin06} K. S. Kuzmin, V. V. Lyubushkin, and V. A. Naumov, Physics of Atomic Nuclei {\bf 69}, 1857 (2006).
\bibitem{Mini-CC-10} A. A. Aguilar-Arevalo et. al (MiniBooNE Collaboration), Phys. Rev. {\bf D} 81, 092005 (2010).
\bibitem{Mini-NC-10} A. A. Aguilar-Arevalo et. al (MiniBooNE Collaboration), Phys. Rev. {\bf D} 82, 092005 (2010).

\bibitem{Paar08} N. Paar, D. Vretenar, T. Marketin and P. Ring, Phys. Rev. {\bf C } 77, 024608 (2008).
\bibitem{Donn72} T. W. Donnelly amd J. D. Walecka, Phys. Rev. {\bf C} 6, 719 (1972).
\bibitem{Wale75} J. D. Walecka, {\it Muon Physics}, edited by V. W. Hughes and C. S. Wu, (Academic Press, New York, 1975).
\bibitem{Weise} T. Ericson and W. Weise, {\it Pions and Nuclei}, (CLARENDON PRESS, Oxford, 1988).
\end{document}